\documentclass[twocolumn]{aastex631}

\usepackage{natbib}
\usepackage{longtable}
\usepackage{rotating}
\usepackage{graphicx}
\usepackage[caption=false]{subfig}
\usepackage{float}
\usepackage{epstopdf}
\usepackage{amsmath}
\usepackage{appendix}
\usepackage{hyperref}
\usepackage{booktabs}
\usepackage{scrextend}
\usepackage{xcolor}
\usepackage{CJKutf8}

\newcommand{\HII}{H{\footnotesize II}}

\newcommand{\NII}{[N~{\footnotesize II}]}

\newcommand{\OI}{[O~{\footnotesize I}]} 
\newcommand{\OIII}{[O~{\footnotesize III}]} 
\newcommand{\SII}{[S~{\footnotesize II}]}

\newcommand{\ha}{H$\alpha$}
\newcommand{\hbeta}{H$\beta$}

\newcommand{\msun}{$M_\odot$}

\begin{document}

\title{{Chandra} and {HST} Observations of Radio-Selected (Wandering) Massive Black Hole Candidates in Dwarf Galaxies}

\author[0000-0003-1055-1888]{Megan R. Sturm}
\affil{eXtreme Gravity Institute, Department of Physics, Montana State University, Bozeman, MT 59717, USA}

\author[0000-0001-7158-614X]{Amy E. Reines}
\affil{eXtreme Gravity Institute, Department of Physics, Montana State University, Bozeman, MT 59717, USA}

\author[0009-0000-9468-7277]{Anne M. Lohfink}
\affil{eXtreme Gravity Institute, Department of Physics, Montana State University, Bozeman, MT 59717, USA}

\author[0000-0003-0573-7733]{Akos Bogdan}
\affil{Center for Astrophysics $\vert$\ Harvard\ \&\ Smithsonian, 60 Garden St., Cambridge, MA 02138, USA}

\author[0000-0002-0765-0511]{Ralph Kraft}
\affil{Center for Astrophysics $\vert$\ Harvard\ \&\ Smithsonian, 60 Garden St., Cambridge, MA 02138, USA}

\author[0000-0003-2686-9241]{Daniel Stern}
\affil{Jet Propulsion Laboratory, California Institute of Technology, 4800 Oak Grove Drive, Pasadena, CA 91109, USA}

\author[0000-0002-7898-7664]{Thomas Connor}
\affil{Center for Astrophysics $\vert$\ Harvard\ \&\ Smithsonian, 60 Garden St., Cambridge, MA 02138, USA}

\author[0000-0003-2511-2060]{Jeremy Darling}
\affil{Center for Astrophysics and Space Astronomy, Department of Astrophysical and Planetary Sciences, University of Colorado, 389 UCB, Boulder, CO 80309-0389,USA;}

\author[0000-0001-8440-3613]{Mallory Molina}
\affil{Vanderbilt University, 2201 West End Ave, Nashville, TN 37235}

\begin{abstract}

We present {Chandra X-ray Observatory} and {Hubble Space Telescope} (HST) follow-up observations of 12 dwarf galaxies from \citet{reines20} that are potential hosts of radio-selected active galactic nuclei (AGNs), eight of which are non-nuclear and possible ``wandering" black holes (BHs). Our multi-wavelength analysis indicates a heterogeneous sample with five radio sources detected at both X-ray and optical wavelengths within positional uncertainties and non-detections for the remaining objects. Of the radio objects detected in the X-ray/optical, three have multi-wavelength evidence for hosting nuclear massive BHs and one object is consistent with an extreme compact starburst. Only one of the off-nuclear radio sources has a significant optical counterpart and we present Palomar spectroscopy that identifies this object as a background AGN. We cannot definitively determine if the seven remaining off-nuclear radio sources are wandering massive BHs in the target dwarf galaxies or background AGNs, although the three sources with the largest offsets have compact radio cores detected with the Very Large Baseline Array and are consistent with expectations for background AGNs \citep{Sargent2022}. Our {HST} sensitivity limits also allow for wandering massive BHs in the target dwarf galaxies that are hosted by stellar clusters with masses $\lesssim 10^6 M_\odot$.

\end{abstract}


\section{Introduction}\label{sec:intro}

We observe supermassive black holes (BHs) in the centers of almost all massive galaxies \citep{magorrian98} across the universe, however, their formation remains a mystery. Multiple potential mechanisms for supermassive BH seeding have been suggested including the collapse of Population III (PopIII) stars \citep{Bond1984, Bromm2004, Fryer2001, Haiman2001}, the collapse of supermassive stars formed from massive gas clouds  \citep{Loeb1994, Begelman2006, Visbal2014, Habouzit2016}, and a gravitational runaway event in a stellar cluster \citep{Miller2002, PortegiesZwart2002}. 

{While BHs with $\sim 10^6-10^7 \: \mathrm{M}_\odot$ have now been observed at high-redshift with {Chandra and} JWST \citep{Maiolino2023a, Bogdan2024, kovacs2024}, directly observing the first seed BHs ($\lesssim 10^5 \: \mathrm{M}_\odot$) remains out of reach \citep{volonterireines2016, schleicher2018, vitoetal2018}. Each seeding model is expected to produce certain detectable signatures in $z \sim 0$ dwarf galaxies; therefore, combining observations of the high-$z$ universe with observations of $z \sim 0$ dwarf galaxies in conjunction with semi-analytic models and simulations may allow us to identify the dominant seeding mechanism. Dwarf galaxies ($M_\star \lesssim 10^{9.5}$ \msun) provide a unique laboratory for studying BH formation and early evolution since they have relatively quiet merger histories compared to massive galaxies, meaning they can provide a `fossil record' of the original BH seeds \citep{volonteri10}. }

For example, models of the BH occupation fraction and BH-galaxy scaling relations differ most substantially in the low-mass regime depending on the seeding mechanism \citep{volonteri08, volonteri09, vanWassenhove10, ricarte18}. Additionally, dwarf galaxies host the smallest known supermassive BHs ($M_{\rm BH} \sim 10^{4-5}$ \msun), which places upper limits on the masses of BH seeds \citep[for a review, see][]{reines22}. Moreover, simulations indicate that as a result of supernova feedback stunting BH growth, dwarf galaxies should host relatively ungrown BHs \citep{angles17, habouzit17}. 

{Since they emit observational signatures across the full electromagnetic spectrum \citep{ho08}, the vast majority of BHs found in dwarf galaxies are accreting and have been identified as active galactic nuclei} (AGNs; e.g., \citealt{reines22, Salehirad2022, pucha2025}). {Radio observations in particular are advantageous since AGNs, regardless of strength, almost always produce radio emission at centimeter wavelengths that is impervious to dust extinction. Additionally, observing high-energy X-ray point-like emission from a galaxy can be indicative of an accreting massive BH \citep{Miller2015, Sacchi2024}. 

However, one complication in the dwarf galaxy regime is that candidate AGNs must be distinguished from potential extraneous sources due to star formation. Objects such as supernova remnants, young supernovae, and \HII \: regions can produce emission in the radio, and some stellar-mass X-ray binaries can reach X-ray luminosities comparable to weakly accreting massive BHs. Since each wavelength regime involves different interloper sources that must be filtered out, multi-wavelength studies are a powerful tool for confirming the presence of AGNs (i.e., objects resembling an AGN at one frequency might not at other frequencies). Combining radio and X-ray observations is particularly useful for finding weakly accreting massive BHs in actively star-forming dwarf galaxies \citep[e.g.,][]{reines11, reines14}. Moreover, high-resolution observations in the visible can help reveal optical counterparts of the radio/X-ray sources and provide details on the host galaxies \citep[e.g.,][]{kimbro21}.}

In this work, we use a combination of X-ray and optical observations to further characterize the sample of radio-selected AGN candidates in dwarf galaxies from \citet{reines20}. \citet{reines20} identify 13 objects that are detected as point-like radio sources at high angular resolution ($\sim 0\farcs25$) with the Karl G.\ Jansky Very Large {Array} and the sources are too luminous in the radio to be explained by normal star formation-related emission. The analysis of \citet{reines20} strongly argues against potential contamination sources including ionized gas from \HII \: regions, supernova remnants (SNRs), younger radio supernovae (SNe), and stellar-mass X-ray binaries (XRBs), supporting the notion that these are truly AGNs.

A particularly unique quality of this specific sample is the fact that the compact radio sources do not always reside in the centers of their dwarf host galaxies. Instead, $\sim 70\%$ of them are offset from their host galaxy centers, with radio-optical positional offsets on the order of a couple of arcseconds ($\sim$ 1-2 kpc). While this phenomenon of ``wandering" black holes is quite unusual in the context of massive galaxies, it has been predicted in multiple simulations for dwarf galaxies hosting lower-mass BHs \citep[$M_{\rm BH} \sim10^{4-6}$ \msun,][]{Bellovary2010, Capelo2015, Tremmel2018, Tremmel2018b, bellovary19, Bellovary2021}. 

Dwarf galaxies generally have low central stellar densities and more irregularly shaped potential wells than their massive counterparts. For a small BH interacting with these kinds of systems, the dynamical friction timescales are much longer than the Hubble time \citep{taffoni2003, Ma2021, Bellovary2021}, meaning that if a BH leaves the center of its host galaxy, it is unlikely to ever return. Consequently, when BH seeds either form outside of a galaxy center \citep{Schneider2002} or, likely more often, when merger events involving tidal stripping \citep{Bellovary2010, Tremmel2015} or gravitational slingshotting \citep{volonteri2005, merritt2004} occur, the result can be a large population of wandering BHs residing in the halos of their host galaxies. \citet{bellovary19} predicts that approximately half of all BHs in dwarf galaxies are wandering, with similar offsets as the \citet{reines20} sample. Despite their prevalence in simulations, wandering BHs remain observationally elusive. Follow-up work verifying the nature of the sample presented by \citet{reines20} is crucial because if the sources are truly AGNs associated with dwarf galaxies, they will provide concrete evidence in support of the theory of wandering BHs.

On the other hand, we cannot rule out the possibility that the radio sources from these putative wandering BHs may in fact be associated with background interlopers (i.e., AGNs) at $z\gg0$. While \citet{reines20} present tentative evidence that the radio sources are associated with the dwarf galaxies rather than background AGNs based on enhanced \OI/\ha\ and \SII/\ha\ line ratios and statistical arguments using known cumulative radio source counts, further corroborating evidence is needed one way or the other. Some hints have come from follow-up Very Long Baseline Array (VLBA) observations. \citet{Sargent2022} observed the 13 radio AGNs associated with dwarf galaxies identified by \citet{reines20} and detected milliarcsecond-scale radio emission towards four of the targets (IDs 2, 28, 48, 65 in \citealt{reines20}) with brightness temperatures consistent with AGNs. However, these objects are the four most distant of the 13 VLA sources from the photocenters of their associated dwarf galaxies (2-4\arcsec), where the chance alignment with a background interloper is higher.

The VLBA nondetections of the other 9 VLA sources do not rule out the presence of AGNs (either in the associated dwarf galaxy or a background interloper); rather, the radio emission is likely resolved out given the dramatically increased angular resolution of the VLBA observations \citep{Sargent2022}. This type of partial resolution is a common phenomenon in radio AGNs \citep{deller2014}. Moreover, we have multi-wavelength confirmation that two of the VLA radio sources not detected with the VLBA are indeed AGNs residing within the target dwarf galaxies. ID 26 is classified as an AGN based on its optical emission line ratios, broad H$\alpha$ emission \citep{reines13}, X-ray luminosity \citep{baldassare17}, and mid-IR colors \citep{latimer21}. ID 82 was confirmed by the AGN coronal line [Fe X], enhanced \OI \: emission at the same location of the radio source and broad H$\alpha$ emission \citep{molina21} \footnote{{While the Si \footnotesize{II} $\lambda$6371 emission line has been misidentified as [Fe X] in some individual objects \citep[e.g.,][]{Herenz2023}, in the case of  ID 82 the other line in the Si \footnotesize{II} doublet (Si \footnotesize{II} $\lambda$6347) is clearly not present and, therefore, there is no reason to doubt the detection of [Fe X].}}.

\begin{deluxetable*}{ccccccccccr}
\tablecaption{Sample of Dwarf Galaxies with Radio Selected AGNs}
\tablewidth{0pt}
\tablehead{
\colhead{ID} & \colhead{NSAID} & \colhead{SDSS Name}  & \colhead{Radio RA} & \colhead{Radio DEC} & \colhead{$N_{\rm H}$} & \colhead{$z$} &\colhead{Dist.} &  \colhead{$M_g$} & \colhead{log($M_\star$/\msun)} \\
\colhead{ }  & \colhead{ } & \colhead{ } & \colhead{(degrees)} & \colhead{(degrees)} & \colhead{($10^{20}$ cm$^{-2}$)} & \colhead{ } & \colhead{(Mpc)} & \colhead{(mag)} & \colhead{ } \\
\colhead{ (1)}  & \colhead{ (2)} & \colhead{ (3)} & \colhead{(4)} & \colhead{(5)} & \colhead{(6)} & \colhead{ (7)} & \colhead{(8)} & \colhead{(9)} & \colhead{(10) } }
\startdata
2          & 26027  & J001900.30+150710.9 & 4.74994 &  15.11973 & 3.57     & 0.03762 &  155 & $-18.42$  & 8.6 \\
6          & 23750  & J010607.17+004633.6  & 16.53045 & 0.77620 & 2.39     & 0.01708 & 70  & $-18.42$  & 9.4 \\
25         & 26634  & J090313.10+482415.5 &135.80403 & 48.40381  & 2.29     & 0.02722 &  112 & $-17.46$  & 8.8 \\
26\tablenotemark{\footnotesize{a}}         & 10779  & J090613.75+561015.5 & 136.55737 &  56.17087 & 2.52     & 0.04697 &  193 & $-18.98$  & 9.4 \\
28         & 12478  & J090908.65+565522.1 &137.28621 & 56.92215  & 3.10     & 0.03147 & 129 & $-17.46$   & 8.3 \\
33         & 16467  & J093138.47+563319.0 & 142.91008&  56.55552 & 2.77     & 0.04940 &  203 & $-16.72$   & 8.3 \\
64         & 66255  & J113648.57+125239.7 & 174.20219&  12.87775 & 2.83     & 0.03404 &  140 & $-18.63$   & 9.3 \\
65         & 101782 & J113642.72+264337.6  &174.17740 & 26.72657 & 2.11     & 0.03306 &  136 & $-18.27$   & 9.2 \\
77         & 3323   & J120058.38$-$034116.3  & 180.24292& -3.68846 & 2.79  & 0.02570 & 106 & $-18.44$   & 9.2 \\
82\tablenotemark{\footnotesize{b}}         & 102751 & J122011.25+302008.1  &185.04694 & 30.33564 & 1.71     & 0.02690 &  110 & $-18.21$  & 9.4 \\
83         & 67389  & J122603.63+081519.0  &186.51518 & 8.25528 & 1.34     & 0.02409 &  99  & $-17.93$   & 9.3 \\
92         & 3602   & J125305.97$-$031258.7 & 193.27487& -3.21632& 1.50     & 0.02213 &  91  & $-19.99$  & 8.6 \\
\enddata
\tablecomments{Column 1: Identification number assigned by \cite{reines20}.
Column 2: Identification number in the NSA.
Column 3: SDSS name.
Column 4: Compact radio source RA from \citet{reines20}.
Column 5: Compact radio source DEC from \citet{reines20}.
Column 6: Galactic neutral hydrogen column density \citep{Bekhti2016}\tablenotemark{c}.  
Column 7: redshift, specifically the \texttt{zdist} parameter from the NSA.
Column 8: Distance to galaxy.
Column 9: absolute $g$-band magnitude corrected for foreground Galactic extinction.
Column 10: log galaxy stellar mass.
The values given in columns 9-12 are from the NSA and we assume $h=0.73$.
\tablenotetext{a}{This object is also an optically-selected AGN with both AGN-like narrow-line ratios and broad H$\alpha$ emission (ID 9 in \citealt{reines13}). It is also a mid-IR-selected AGN (ID 2 in \citealt{latimer21}).}
\tablenotetext{b}{This object also has optical AGN signatures, including the coronal line [Fe X], enhanced [O I] emission
at the same location of the radio source, and broad H$\alpha$
emission \citep{molina21}.}
\tablenotetext{c}{Retrieved via \url{https://heasarc.gsfc.nasa.gov/cgi-bin/Tools/w3nh/w3nh.pl}.}
}
\label{tab:sample}
\end{deluxetable*}

{Here we continue our follow-up campaign of 12 objects from the \citet{reines20} sample of radio AGNs associated with dwarf galaxies.} Our paper is primarily focused on new X-ray observations from the {Chandra X-ray Observatory} (Section \ref{sec:xray}) and optical imaging from the {Hubble Space Telescope} ({HST}, Section \ref{sec:optical}). {In Section \ref{sec:spectra_ID64}, we present the Palomar spectrum of one of our targets. We combine the pre-existing radio detections with our new X-ray, and optical detections/limits into a multi-wavelength view of this sample in Section \ref{sec:multiwavelength}. In this section, we also address the implications of our X-ray/optical non-detections and provide a detailed, multi-wavelength analysis of each object individually. We use a flat $\Lambda$CDM cosmology with {$\Omega_{\rm m}$} = 0.3 and H$_0$ = {73} km s$^{-1}$ Mpc$^{-1}$.}

\section{Sample of Dwarf Galaxies with Radio-Selected AGNs} 
\label{sec:sample}

The dwarf galaxies studied in this work are associated with {candidate} AGNs detected via luminous compact radio emission by \citet{reines20}. In their study, \cite{reines20} start with the full NASA-Sloan Atlas (v0\_1\_2)\footnote{\url{http://nsatlas.org/}}, which is a catalog of SDSS galaxies with redshifts $z<$0.055. They made cuts on stellar mass and absolute magnitude, requiring $\mathrm{M}_\star \leq 3 \times 10^{9}$~\msun \! ($\sim$the mass of the Large Magellanic Cloud) and $\mathrm{M}_g$ and $\mathrm{M}_r >$ -20, which resulted in a parent sample of 43,707 dwarf galaxies. They cross-match this parent sample of dwarf galaxies with the VLA Faint Images of the Radio Sky at Twenty-centimeters (FIRST) Survey \citep{becker95} catalog (version 2013 Jun 5). This resulted in 148 matches, which were targeted for further observations with the VLA at higher resolution ($\sim$0\farcs25) and sensitivity ($15 \mu$Jy). Ultimately, 111 of the targets were observed due to scheduling priorities. The new VLA observations have $\sim 20\times$ better angular resolution and $10\times$ the sensitivity of FIRST. \citet{reines20} detected compact radio sources towards 28 confirmed dwarf galaxies, 13 of which were compelling massive BH candidates based on radio luminosities exceeding that expected from star-formation-related emission.
 
This paper is focused on 12 of these objects, for which we present {Chandra} and {HST} observations. We adopt the ID numbers used in \citet{reines20}. Four of the dwarf galaxies have radio sources that are centrally located (IDs 26, 82, 83 and 92), while 8 radio sources are offset from the nuclei of the target galaxies (IDs 2, 6, 25, 28, 33, 64, 65 and 77). Properties of the 12 dwarf galaxies in our sample are summarized in Table \ref{tab:sample}. One galaxy from \citet{reines20}, ID 48, is excluded from our follow-up observations since it lacks an SDSS spectrum and the radio source has a faint red optical point-source counterpart in DECaLS \citep[Dark Energy Camera Legacy Survey;][]{dey19} imaging, indicative that it is likely a background quasar. Moreover, \citet{Sargent2022} also detect the source with the VLBA and note that its distance from the center of the dwarf galaxy is consistent with expectations for a background AGN.

\begin{deluxetable}{clccc}
\tabletypesize{\footnotesize}
\tablecaption{\textit{Chandra} X-ray Observations}
\tablewidth{0pt}
\tablehead{
\colhead{ID}  & \colhead{Date observed}  & \colhead{Obs ID} & \colhead{Exp.\ time (ks)}  }
\startdata
2       & 2022 Sep 28 & 25269 & 16.84  \\
2       & 2022 Oct 18 & 26031 & 17.13  \\ 
6       & 2022 Jun 12 & 25270 & 7.95   \\
{6}   & {2001 Sep 18} & {2180}  & 3.76     \\
25      & 2022 Jan 26 & 25271 & 16.73 \\
26      & 2014 Dec 26 & 17033 & 15.83  \\
28      & 2022 May 28 & 25272 & 18.12  \\
33      & 2022 Sep 01 & 25273 & 17.89  \\
33      & 2021 Dec 08 & 26032 & 20.76  \\ 
64      & 2021 Nov 07 & 24735 & 13.92  \\
64      & 2021 Nov 08 & 26181 & 13.93  \\
65      & 2022 May 15 & 26033 & 9.93   \\
65      & 2023 Jul 05 & 27940 & 8.28   \\
65      & 2023 Jul 05 & 25274 & 9.92   \\
77      & 2021 Apr 15 & 24736 & 17.88  \\
82      & 2022 Jul 31 & 25275 & 17.82  \\
83      & 2022 Feb 27 & 25276 & 14.86  \\ 
92      & 2019 Mar 06 & 21518 & 5.01   \\
\enddata
\tablecomments{ New \textit{Chandra} observations were taken for all galaxies except IDs 26 and 92, for which we use archival data.
}
\label{tab:cxo}
\end{deluxetable}

\section{Chandra X-ray Observations}
\label{sec:xray}

X-ray observations of our target galaxies were taken with {Chandra} between 2014 December 26 and 2023 March 13, with exposure times ranging between 3.8 and 20.8 ks. We use archival {Chandra} data for two of our galaxies (IDs 26 and 92) and obtain new observations for our remaining 10 galaxies. In each observation the target galaxy was placed at the center of the ACIS-S3 chip, except for the archival observation of ID 92 (OBSID 21518) in which the galaxy was placed at the aimpoint of the ACIS-I3 chip. A summary of the {Chandra} observations is given in Table \ref{tab:cxo} and the observations can be found at 
\dataset[10.25574/cdc.367]{https://doi.org/10.25574/cdc.367}.

\subsection{Chandra Data Reduction and Astrometry}
\label{sec:chandra_obs}

We first reprocessed the data using \texttt{CIAO} software v4.16 \citep{fruscione06} and applied calibration files (CALDB 4.11.5) to create new level 2 event files. 

We then improve the {Chandra} absolute astrometry, which is typically good to $\sim$ 0\farcs4. To achieve this, we began by creating an image in the broad ACIS energy band (0.5-7\,keV) for each observation. Following this, we generated a list of detected X-ray point sources using the \texttt{CIAO} function \texttt{wavdetect}, excluding any sources that are detected with a significance of less than 3$\sigma$ above the background or falling within 5\arcsec of the target galaxy. The detected X-ray sources were matched to point sources in the Gaia DR3 catalog using the \texttt{CIAO} function \texttt{wcs\_match}. We found several matching \texttt{wavdetect} and Gaia sources for each observation, and updated the {Chandra} astrometry using a translation correction. The resulting geometric shifts are generally small ($\leq$ 2 pixels) with the exception of OBSIDs 25271 and 25273 where unphysically large corrections were suggested by the software. Consequently, these two observations were not aspect corrected, however, they are also both non-detections with no possible targets within 5\arcsec \: of the target galaxies, making the lack of precision astrometry affect our results minimally, at most.

\begin{deluxetable*}{cccccrrrr}
\tabletypesize{\footnotesize}
\tablecaption{X-ray Photometry}
\tablewidth{0pt}
\tablehead{
\colhead{ID} & \colhead{R.A.} & \colhead{Decl.} & \colhead{R.A./Decl. error} & \colhead{Net Rate ($10^{-4}$ cts/s)} & \multicolumn{2}{c}{Flux ($10^{-15}$ erg s$^{-1}$ cm$^{-2}$)} & \multicolumn{2}{c}{Luminosity (log(erg s$^{-1}$))} \\
 \cmidrule(l){6-7} \cmidrule(l){8-9}
\colhead{ } & \colhead{(deg)} & \colhead{(deg)} & \colhead{($10^{-5}$\,deg)} & \colhead{0.5-10 keV} & \colhead{0.5-2 keV} & \colhead{2-10 keV} & \colhead{0.5-2 keV} & \colhead{2-10 keV} \\ 
\colhead{ (1)}  & \colhead{ (2)} & \colhead{ (3)} & \colhead{(4)} & \colhead{(5)} & \colhead{(6)} & \colhead{ (7)} & \colhead{(8)} & \colhead{(9)}}
\startdata
2\tablenotemark{\footnotesize{a}}         & \nodata    & \nodata   & \nodata  & $<$0.79 & $<$1.02 & $<$1.97  & $<$39.5    & $<$39.8   \\
6\tablenotemark{\footnotesize{a}}         & \nodata    & \nodata   & \nodata  & $<$2.3 & $<$1.67 & $<$5.57 & $<$39.0    & $<$39.5  \\
25        & \nodata    & \nodata   & \nodata  & $<$1.4 & $<$1.74 & $<$3.19     & $<$39.4    & $<$39.7   \\
26  	  & 136.557593 & 56.170764 & 5.3/3.4 & $8.7_{-3.4}^{+4.5}$  &  4.40$_{-1.80}^{+2.51}$    & 2.51$_{-2.14}^{+4.80}$   & 40.3$\pm 0.2$    & 40.1$_{-0.9}^{+0.4}$   \\
28        & \nodata    & \nodata   & \nodata  & $<$1.3 & $<$1.56 & $<$2.96  & $<$39.5    & $<$39.8   \\
33\tablenotemark{\footnotesize{a}}        & \nodata    & \nodata   & \nodata  & $<$1.0 & $<$0.86 & $<$0.27    & $<$39.6    & $<$39.1   \\
64\tablenotemark{\footnotesize{a}}        & 174.202138 & 12.878138 & 0.8/0.9    & 222.0$\pm 16.0$    & 115.0$_{-12.0}^{+13.0}$   & 305.0$_{-30.0}^{+29.0}$ & 41.4$\pm 0.1$\tablenotemark{\footnotesize{b}}    & 41.9$\pm 0.1$\tablenotemark{\footnotesize{b}}     \\
65\tablenotemark{\footnotesize{a}}        & \nodata    & \nodata   & \nodata   & $<$3.3    & $<$2.88 & $<$5.32   & $<$39.8    & $<$40.1  \\
77  	  & \nodata    & \nodata   & \nodata  & $<$1.3 & $<$1.52 & $<$2.96  & $<$39.3    & $<$39.6   \\
82        & 185.047155     & 30.335761   & 10.0/7.6  & 2.1$_{-1.4}^{+2.4}$ & $<$2.76 & 3.69$_{-2.96}^{+4.92}$   & $<$39.6    & 39.7$_{-0.3}^{+0.4}$   \\
83 	      & 186.514998 & 8.255182  & 1.3/1.2 &  49.0$_{-9.4}^{+9.5}$  & 36.4$_{-8.5}^{+9.9}$ & 46.6$_{-12.7}^{+15.6}$   & 40.6$\pm 0.1$     & 40.7$_{-0.1}^{+0.2}$   \\
92  	  & 193.274734 & $-$3.216360 & 4.7/4.1 & 23.2$^{+13.4}_{-9.7}$ & 17.6$_{-9.1}^{+14.1}$  & $24.6_{-14.7}^{+24.0}$    & 40.2$\pm 0.3$  & 40.4$_{-0.4}^{+0.2}$   \\
\enddata
\tablecomments{Column 1: Identification number assigned by \cite{reines20}. 
Column 2: right ascension of X-ray source.  
Column 3: declination of X-ray source. 
Column 4: uncertainties of X-ray coordinates.
Column 5: net count rate after applying an aperture correction.  Error bars represent 90\% confidence intervals.
Columns 6-7: fluxes corrected for Galactic absorption; calculated using a photon index of $\Gamma = 1.8$.
Columns 8-9: photometric log luminosities corrected for Galactic absorption; calculated using a photon index of $\Gamma = 1.8$ and assuming the X-ray source is in the target dwarf galaxy. 
Log luminosities in the soft/hard band but calculated from the spectral modeling are: $40.9\pm 0.2$/$39.0\pm 0.4$ (ID 26), $41.47 \pm 0.03$/$41.86 \pm 0.03$ (ID 64), $40.7\pm 0.1$/$40.8 \pm 0.1$ (ID 83), $40.4\pm 0.2$/$40.8\pm 0.2$ (ID 92).
}

\tablenotetext{a}{\label{merged}These galaxies had multiple \textit{Chandra} observations that required merging prior to analysis. Any coordinates listed are from the merged images; the merged count rate and fluxes are reported here (see Section \ref{sec:xrayemis}).} 
\tablenotetext{b}{{If the X-ray emission originates from the background AGN at z=0.761 rather than the target dwarf galaxy (see Section \ref{sec:spectra_ID64}), these luminosities would be 44.4 erg s$^{-1}$ in the 0.5-2 keV band and 44.9 erg s$^{-1}$ in the 2-10 keV band.}}
\label{tab:xray}
\end{deluxetable*}

To find X-ray candidate sources, we re-run \texttt{wavdetect} after the astrometry correction unless multiple observations exist for the same target. {Five} of our target galaxies had multiple {Chandra} observations: ID 2 (OBSIDs 25269 and 26031), {ID 6 (OBSIDs 2180 and 25270)}, ID 33 (OBSIDs 25273 and 26032), ID 64 (OBSIDs 24735 and 26181), and ID 65 (OBSIDs 25274, 26033, and 27940). These observations are merged prior to the final imaging analysis. We first merge the individually astrometrically corrected observations using the \texttt{CIAO} function \texttt{merge\_obs}. Then, we run \texttt{wavdetect} on the merged images, similarly to our treatment of the individual images.

\begin{figure}
    \centering
    \includegraphics[width=8.5cm]{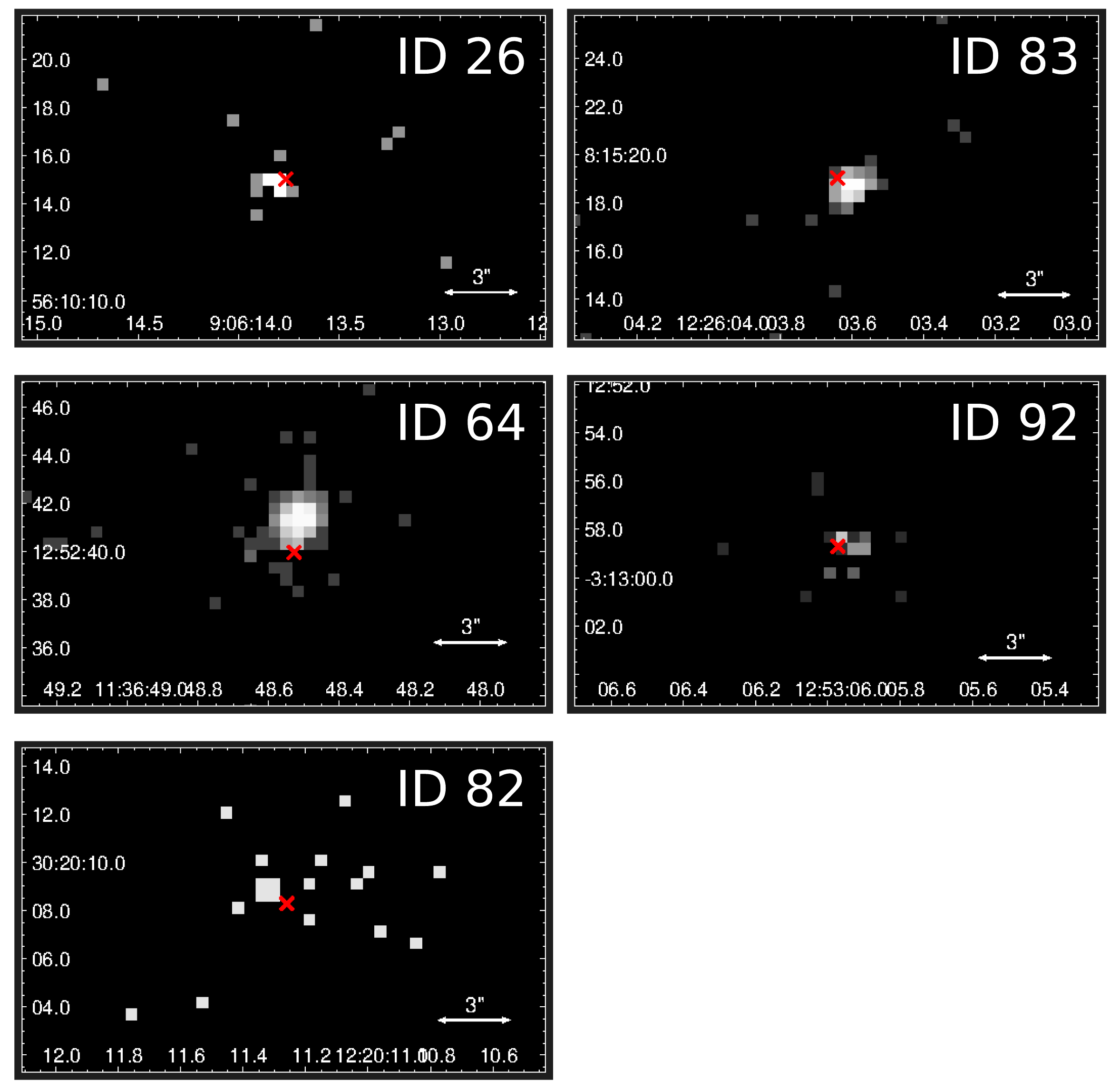}
    \caption{{Chandra 0.5-7 keV images for our targets galaxies with X-ray detections. The red cross shows the locations of the detected compact radio sources. While the centroids of the Chandra sources do not always directly align, the sources are roughly positionally consistent.}}
    \label{fig:xray_detections}
\end{figure}

We restrict further analysis to $>3\sigma$ detected sources within 5\arcsec \: (i.e., a similar radius to the cross-matching radius used by \citealt{reines20}). Within such a radius at the depth of our X-ray observation, the chance of a background source is very low \citep{moretti03} \footnote{{We also check if any of our sources are co-spatial with a source in the MILLIQUAS catalog \citep{Flesch2023} and do not find any such matches.}}. {This reveals candidate sources (distance to radio position $<$ 5\arcsec)} in IDs 26, 82, 83 and 92 (Figure \ref{fig:xray_detections}). {The detected X-ray sources are consistent with being central to the host galaxy, with the exception of ID 64 (likely a background AGN).} Of the five galaxies requiring merging, one (ID 64) had detected sources in the galaxy region (this holds true in both the merged analysis and the analysis of the individual observations) at a significance of 3$\sigma$. We note here, that in ID 65, one observation had a tentative X-ray source detection (OBSID 27940) below our 3$\sigma$ significance threshold; no source or hint of a source was detected in the other two observations of ID 65 (despite one been taken on the same day at similar depth). 

Upon the successful detection of the X-ray source in ID 64, we checked the astrometric alignment of the unmerged images by comparing the centroids of the detected source in the unmerged OBSID event files. In ID 64, the astrometric offset between the two centroids was $\ll0\farcs492$ (one {Chandra} pixel), negating the need for further fine astrometric correction (see Figure \ref{fig:xray_detections}).  As such, no source centroids could be found, so no further astrometric corrections were applied.

\subsection{X-ray Photometry}
\label{sec:xrayemis}

Our X-ray photometry builds upon the initial imaging results. In the case of a candidate source detection using \texttt{wavdetect}, we center our photometry on the coordinates identified by \texttt{wavdetect}. If no source is detected, we revert to using the original coordinates of the radio source for photometry.
Source counts are extracted using the \texttt{CIAO} function \texttt{srcflux}, with circular apertures corresponding to the $90\%$ enclosed energy fraction at 1.0 keV. {To estimate the background counts, we use circular annuli centered on the source region, with an inner radius equal to 1.7$\times$ the radius of the source aperture and an outer radius set at $3\times$ the source aperture radius. }

The \texttt{srcflux} function is also used to calculate unabsorbed X-ray fluxes in the hard and soft energy bands. We adopt a power-law spectral model with a photon index $\Gamma = 1.8$, {which is typical for both ultraluminous X-ray sources at these energies }\citep{swartz08} and low-luminosity AGNs \citep{ho08, ho09}. Galactic column densities are obtained from the maps of \cite{Bekhti2016} and applied via the \texttt{TBabs} model \citep{Wilms2000}. In case of multiple observations, \texttt{srcflux} is run on the stack of all event files and the merged source fluxes are reported. If no source was detected, the \texttt{srcflux} returns estimates for the 90\,\% confidence upper limits of net count rate and fluxes. {The inferred 0.5–10~keV flux depends on the assumed spectral slope; varying $\Gamma$ between 1.3 and 2.3 changes the flux estimate by roughly a factor of $\sim$2.5 (0.4 dex), which represents the systematic uncertainty associated with the spectral assumption.}

The net count rates, estimated unabsorbed fluxes, and luminosities are summarized in Table \ref{tab:xray}. The 2-10 keV luminosities range from log$(L_{\rm 2-10keV}/{\rm erg\,s}^{-1}) = 39.7$-$41.9$, assuming the sources reside within the target dwarf galaxies (see Table \ref{tab:xray}). These luminosity values should be considered as lower limits, as we do not account for potential intrinsic absorption by the sources. For observations where no X-ray sources are detected, we use estimated minimum detectable fluxes to establish upper limits on X-ray source luminosities.

\subsection{X-ray Sources}
\label{sec:xraysrc}
\subsubsection{Spectral Information}
Spectroscopy provides deeper insights into the nature of the detected sources compared to photometry, and thus we created spectra for each of the five detected sources (see Figure~\ref{fig:xrayspectra} in Appendix~\ref{sec:appendix}). We extract spectra using the \texttt{CIAO} task \texttt{specextract} using a circular source region centered on the source coordinates, while the background region is an annular area surrounding the source region. In the case of ID 64, the spectra obtained from the individual observations were combined to create a summed spectrum with corresponding responses using the task \texttt{combine\_spectra}. All resulting spectra were then binned prior to modeling using \texttt{grppha} so that they possess a minimum of 1 count per bin (or 20 counts per bin in the case of ID 64). The spectra are modeled in the 0.5-10\,keV range. As the spectrum of ID 64 is the only one with sufficient quality, it is modeled using $\chi^2$ fitting while all other spectra are modeled using Cash statistics. Whenever possible, the absorption corrected fluxes were found using the \texttt{cflux} model, and the corresponding luminosities can be found in Table~\ref{tab:xray}. In the following, we briefly describe our X-ray spectral findings for each of the detected sources. All our spectral modeling was performed in \texttt{XSPEC} with spectral parameter uncertainties determined at the 90\,\% confidence level, using the error command. Galactic absorption is again described by the \texttt{TBabs} model.

\textbf{ID 26:} The spectral quality of the spectrum of ID 26 is poor especially at higher energies. The X-ray spectrum is very steep ($\Gamma=5.1\pm 1.4$) for a fit with a power law modified by Galactic absorption ($N_H$ fixed). Although the spectral quality is poor, the spectrum is not quite consistent with a power law attenuated by Galactic absorption (best fit C-Stat$=17.8$/13 dof), and visible residuals remain in the 0.8-1.0\,keV range. If we replace the power law with the spectrum of a diffuse collisionally-ionized gas with solar abundances (\texttt{XSPEC} model \texttt{apec}), the residuals almost entirely vanish and the fit improves to C-Stat$=10.9$/13 dof. The temperature of the plasma is found to be 0.5$\pm$0.3\,keV, which is in line with what is commonly found for star-forming galaxies, see for example \citet{Kyritsis2025} for a recent example. As an alternative scenario, we attempted to free the column density in the absorbed power law scenario during the modeling to see whether the X-ray source could also be a slightly enshrouded, young, highly accreting, black hole but we found that such a model was not able to reproduce the 0.8-1.0\,keV feature as well as the collisionally ionized plasma model. 

\textbf{ID 64:} The summed X-ray spectrum of this target is consistent with a power law modified by Galactic absorption (best fit $\chi^2=25.5$/28 dof). Calculating the 1$\sigma$ uncertainties in \texttt{XSPEC} we find the source to possess a fairly hard spectrum ($\Gamma=1.51\pm 0.13$). There are no indications of any other components in the spectrum nor is intrinsic absorption indicated. We obtain the 2-10\,keV flux from the power law component via the cflux model, finding $\log(f_{2-10\,\mathrm{keV}}/\mathrm{erg}\,\mathrm{s}^{-1}\,\mathrm{cm}^{-2})=-12.51 \pm 0.05$.

As we discuss in section \ref{sec:individual objects},  it is quite likely that this source is in fact a  background AGN at redshift $0.761$. If true, then the modeled  0.5-10\,keV band would correspond to an actual intrinsic energy range of $\sim 0.9-17.6$\,keV. Within this energy band, intrinsic absorption cannot be ruled out at the same level of confidence, however absorption in excess of $N_H=10^{22}$cm$^{-2}$ is not consistent with the observed spectrum. Such as background AGN would have a $2-10$\,keV luminosity of approximately $7\times 10^{44}$\,erg/s. 

\textbf{ID 82:}
ID 82 has low photon count and a usable spectrum can therefore not be extracted. Nevertheless, we have some spectral information in that the target is only detected at hard energies. This could be suggestive of an obscured accreting black hole. Using the \texttt{WebPIMMS} v4.51 tool to estimate how much absorption is consistent with the observed fluxes, we find that a column of a few $\times 10^{21}$\,cm$^{-2}$ is consistent with our findings, assuming the spectrum can be roughly described as an absorbed power law with a photon index in the range of 1.8-2.2.

\textbf{ID 83:}
The spectrum of ID 83 is statistically consistent with a power law modified by Galactic absorption (C-Stat$=58.9$/61 dof). The photon index derived from such a description of the data is 2.0$\pm$0.4. 

\textbf{ID 92:}
Due to the extremely short exposure time of only 5\,ks, the spectrum of ID 92 is very poor. It is very well described by a power law modified by Galactic absorption (C-stat$=11.9/16$). The constraining power of the spectrum is minimal and therefore the power law photon index is poorly constrained to 1.4$\pm 0.7$.

\subsubsection{Expected Contribution from Ultraluminous X-ray Sources} \label{sec:xrbs}

Our high-resolution Chandra observations reveal X-ray sources toward five of our target dwarf galaxies with luminosities of $L_{\rm 2-10~keV} \sim 10^{40-42}$ erg s$^{-1}$. If the X-ray and radio emission originate from the same object, the high {\it radio} luminosities provide strong evidence for AGNs as the origin, as stellar-mass objects have much lower ratios of radio to X-ray emission. However, if the radio and X-ray emission originate from separate objects, which is possible given the astrometric uncertainties, it is possible that ultraluminous X-ray sources (ULXs), generally defined to be off-nuclear X-ray sources with $L_{\rm X} > 10^{39}$ erg s$^{-1}$ \citep{kaaret17}, could be responsible for the observed X-ray luminosities. Therefore, we assess the likelihood of ULXs here using the  relation in \citet{lehmer21} between the number of ULXs expected in a galaxy based on the star formation rate (SFR) and gas-phase metallicity. 

We estimate SFRs for our target dwarf galaxies using far-UV (FUV; 1528 \AA) and mid-infrared (IR; 25 $\mu$m) luminosities:

\begin{equation} \label{eq:sfrs}
\begin{gathered}
 \textrm{log SFR}(\textrm{\msun~yr}^{-1}) = \textrm{log }L\textrm{(FUV)}_{\textrm{corr}} - 43.35 \\
 L \textrm{(FUV)}_{\textrm{corr}} = L\textrm{(FUV)}_{\textrm{obs}} + 3.89 L\textrm{(25 $\mu$m)}
\end{gathered}
\end{equation}

\noindent
\citep{hao11, kennicutt12}, where the uncertainty for the estimated SFR is ${\sim}0.13$ dex.  We retrieve FUV magnitudes from the \textit{Galaxy Evolution Explorer} (\textit{GALEX}), and for the IR data use 22 $\mu$m magnitudes from the \textit{Wide-field Infrared Survey Explorer} (\textit{WISE}) instead of 25 $\mu$m magnitudes from the \textit{Infrared Astronomical Satellite} (\textit{IRAS}). None of our galaxies are detected by \textit{IRAS}, while all twelve have \textit{WISE} detections, and the ratio between 25 $\mu$m and 22 $\mu$m flux densities is expected to be ${\sim}1$ \citep{jarrett13}. As discussed in \citet{latimer21}, ID 26 is a mid-IR-selected AGN, implying the UV$+$IR SFR is likely severely overestimated due to AGN emission at mid-IR wavelengths. Therefore, we adopt the uncorrected UV SFR of 0.14 $M_\odot$/yr from that work. The resulting SFRs for the five galaxies with X-ray detections are 0.14, 0.34, 0.29, 0.16 and 10.49 $M_\odot$ yr$^{-1}$ for IDs 26, 64, 82, 83, and 92, respectively.

Following \citet{latimer21}, we estimate the metallicities of our dwarf galaxies based on the ratio of \NII / H$\alpha$ as calculated in \cite{reines15} and the relation given in \citet{Pettini2004}. The metallicities of the galaxies, excluding ID 92, range from 12 $+$ log(O/H) $=$ 8.46-8.66 with a mean of 8.55. {We are unable to include ID 92 in this calculation due to the poor quality of the SDSS spectrum. From the relation derived in \citet[][see their Figure 6]{lehmer21} we expect $\sim$0.1 and ULX with $L_{\rm X} > 10^{40}$ erg s$^{-1}$ per unit SFR at a metallicity of 12 $+$ log(O/H) $=$8.5. Given the total SFR for these four galaxies is 0.93 \msun/yr, the expected number of ULXs is $\lesssim 0.1$. Therefore, it is likely that the detected X-ray sources in these four galaxies are due to AGNs.}

{ID 92 is a blue compact dwarf galaxy and is in the sample studied by \citet{thuan05}. They find 12 $+$ log(O/H) $=$ 8.05 and we expect $\sim$0.2 ULX with $L_{\rm X} > 10^{40}$ erg s$^{-1}$ per unit SFR at this lower metallcity \citep{lehmer21}. Therefore, ID 92 may have 2 ULXs given its relatively high SFR and the presence of a massive star forming region seen in HST imaging (see Section \ref{sec:individual objects}).} 



\begin{deluxetable}{cccccc}
\tabletypesize{\footnotesize}
\tablecaption{\textit{HST} Observations}
\tablewidth{0pt}
\tablehead{
\colhead{ID} & \colhead{Program ID} & \colhead{Date observed} & \multicolumn{3}{c}{Exp.\ time (s) in each filter} \\
\hline
\colhead{ } & \colhead{ } & \colhead{} & \colhead{F475W} &  \colhead{F680N} & \colhead{F814W} }
\startdata
2 & 16843 & 2021 Dec 05 & 584 & 1060 & 600 \\
25 & 16843 & 2022 Jan 09 & 600 & 1172 & 600 \\
28 & 16843 & 2021 Dec 31 & 636 & 1120 & 720 \\
33 & 16843 & 2022 Feb 13 & 626 & 1120 & 720 \\
65 & 16843 & 2021 Dec 30 & 584 & 1060 & 608 \\
82 & 16843 & 2021 Dec 27 & 584 & 1060 & 624 \\
83 & 16843 & 2022 Jan 02 & 532 & 1030 & 570 \\
\hline
& & & \colhead{F475W} &  \colhead{F665N} & \colhead{F814W} \\
\hline
6 & 16843 & 2021 Dec 27 & 574 & 1060 & 600 \\
\hline
& & & \colhead{F438W} &  \colhead{F680N} & \colhead{F814W} \\
\hline
64 & 16661 & 2022 Apr 25 & 810 & 2391 & 996 \\
77 & 16661 & 2021 Nov 23 & 810 & 2385 & 990 \\
92 & 16661 & 2022 Jan 06 & 810 & 2382 & 987 \\
\hline
& & & \colhead{F275W} &  \colhead{F606W} & \colhead{F110W} \\
\hline
26 & 13943 & 2015 Feb 18 & 900 & 648 & 514 \\
\enddata
\tablecomments{New \textit{HST} observations were taken for all of the galaxies in our sample except ID 26, for which we use archival \textit{HST} imaging. {Our HST observations can be found at \dataset[10.17909/h9mh-tg88]{http://dx.doi.org/10.17909/h9mh-tg88}.}
}
\label{tab:hst}
\end{deluxetable}

\section{HST Optical Observations}
\label{sec:optical}

Our new {HST} optical images were taken between 23 November 2021 and 25 April 2022 in two programs (program IDs 16661 and 16843) using the Wide Field Camera 3 (WFC3) UVIS imaging instrument. We observed each galaxy for one orbit in three filters. We used a narrow-band \ha\  filter (either F665N or F680N, depending on the redshift of the target galaxy) and two wide-band filters including F814W (centered at 8039\AA) and either F438W or F475W (centered at 4326\AA \: or 4773\AA, respectively). 

For program ID 16843, each galaxy was observed in one orbit with exposure times of $\sim$17-20 minutes in the narrow-band \ha \: filter and $\sim$9-12 minutes in the wide-band filters. For program ID 16661, each galaxy was observed over two orbits for $\sim$40 minutes in the narrow-band \ha\ filter and $\sim$13-17 minutes in the wide-band filters. For ID 26 we use archival data from program ID 13943 (PI Reines) from 18 February 2015. The galaxy was observed in three wide-band filters: for 15 minutes in the F275W filter (centered at 2710\AA), for 10 minutes in the F606W filter (centered at 5889\AA) and for 8.5 minutes in the IR F110W filter (centered at 1153\AA). {A summary of the {HST} observations is given in Table \ref{tab:hst} and the observations can be found at \dataset[10.17909/h9mh-tg88]{http://dx.doi.org/10.17909/h9mh-tg88}.}

\subsection{HST Data Reduction and Astrometry}

We obtained the calibrated and drizzled Hubble Advanced Product (HAP) Single Visit Mosaics (SVMs) produced
by the STScI data reduction pipeline. The HAP SVMs are drizzled onto the same north-up pixel grid and are aligned to a common astrometric reference frame. The pipeline attempts to improve both the relative astrometry between filters as well as the absolute astrometry by aligning sources in the images to the \textit{Gaia} source catalog.

\begin{deluxetable*}{ccccccccccccccccc}
\tablecaption{Optical Photometry \label{tab:optical_mags}}
\tablehead{
\colhead{ID} &  \colhead{F275W}  &   \colhead{F438W} & \colhead{F475W} & \colhead{F606W} & \colhead{F665N} & \colhead{F680N} & \colhead{F814W} & \colhead{F110W} }
\startdata
2 &    \nodata  &  \nodata    &   $>$24.34    &  \nodata    &  \nodata    &   $>$24.50    &   $>$24.79  &\nodata  \\
6 & \nodata &  \nodata    &   $>$23.71    &  \nodata    &   $>$23.83    &  \nodata    &   $>$24.08    &\nodata \\
25 &    \nodata  &  \nodata    &   $>$23.29    &  \nodata    &  \nodata    &   $>$23.53    &  $>$23.73   &\nodata   \\
26 &    20.23    &  \nodata    &  \nodata    &  20.03     &  \nodata    &  \nodata    &  \nodata & 21.28   \\
28 &   \nodata &  \nodata    &  $>$24.04    &  \nodata    &  \nodata    &  $>$24.28    &  $>$24.50   &\nodata   \\
33 &     \nodata  &  \nodata    &  $>$24.22     &  \nodata    &  \nodata    &  $>$24.13     &   $>$24.62   &\nodata  \\
64 &    \nodata  &  21.19     &  \nodata    &  \nodata    &  \nodata    &  21.74     &  21.86   &\nodata   \\
65 &   \nodata  &  \nodata    &   $>$24.03    &  \nodata    &  \nodata    &   $>$24.10    &   $>$24.22  &\nodata   \\
77 & \nodata  &  $>$22.84    &  \nodata    &  \nodata    &  \nodata    &   $>$23.03     &   $>$23.28  &\nodata   \\
82 &    \nodata  &  \nodata    &  20.30     &  \nodata    &  \nodata    &  20.19     &  21.12   &\nodata   \\
83 &  \nodata  &  \nodata    &  20.42     &  \nodata    &  \nodata    &  20.20     &  20.42   &\nodata   \\
92 &    \nodata  &  17.24     &  \nodata    &  \nodata    &  \nodata    &  16.74     &  18.76   &\nodata   \\
\enddata
\tablecomments{Optical ST magnitudes in each filter measured from photometry at the location of the compact radio source for galaxies in our sample. If the flux within an aperture with radius 0\farcs2 exceeds the background flux by 3$\sigma$, we report the detected magnitude. In the case that the flux does not exceed 3$\sigma$ of the background we report a lower limit on the magnitude (upper limit on the brightness) of the non-detection.  {An ellipses indicates that the target was not observed with that specific filter.}}
\label{tab:optical_mags}
\end{deluxetable*}

When necessary, we made additional fine adjustments to the astrometry in each of our images by comparing common sources with DECaLS, which has absolute astrometry accurate within $\sim 0\farcs1$. We shifted each SVM image individually to match the DECaLS images and then updated the WCS for our images accordingly. {The resulting astrometric corrections had a maximum RA shift of 0\farcs39 and a maximum DEC shift of 0\farcs25.}

\subsection{Optical Photometry}
\label{sec:optical_photometry}

For a given galaxy, we perform aperture photometry in the regions around the detected radio source to determine if there is detectable emission (i.e., an optical counterpart) in each band. {We search for optical counterparts located within 0\farcs25 of the radio source, which is equivalent to the resolution of the VLA observations presented in \citet{reines20}. This search radius also allows for uncertainties in the absolute astrometry between the VLA and {HST} observations ($\sim 0\farcs1$ each). We perform aperture photometry on identifiable sources using a source aperture of 0\farcs2 centered on the optical source and a nominal background annulus with an inner radius of $r_{in}=0\farcs3$ and outer radius of $r_{out}=0\farcs4$. {Since the uncertainty associated with this measurement is dominated by the background, we vary the background annulus and find an average error of $\sim$6\%.} Aperture corrections were applied to our final background-subtracted flux densities using the appropriate encircled energy fractions for a 0\farcs2 aperture. To ensure solid detections, we also require the background-subtracted source flux density to be larger than $3\sigma$ above the median background. In cases where there is no obvious optical counterpart, we derive upper limits by performing aperture photometry using the same method described above but with the aperture centered on the radio source. The measured values and upper limits are given in Table \ref{tab:optical_mags}}.

We find five galaxies (IDs 26, 64, 82, 83 and 92) that have significant optical detections at the location of the radio source in all of their filters. {These same five galaxies also have overlapping {Chandra} X-ray detections (see Figure \ref{fig:rgb} and Table \ref{tab:summary})}. Four of the radio sources with optical detections are located in the nuclei of their host dwarf galaxies, while the remaining radio source in ID 64 is off-nuclear and our Palomar spectroscopy of the relatively bright optical counterpart reveals that it is a background AGN (see Section \ref{sec:spectra_ID64}).

\begin{figure*}
    \centering
    \includegraphics[width=6.0in]{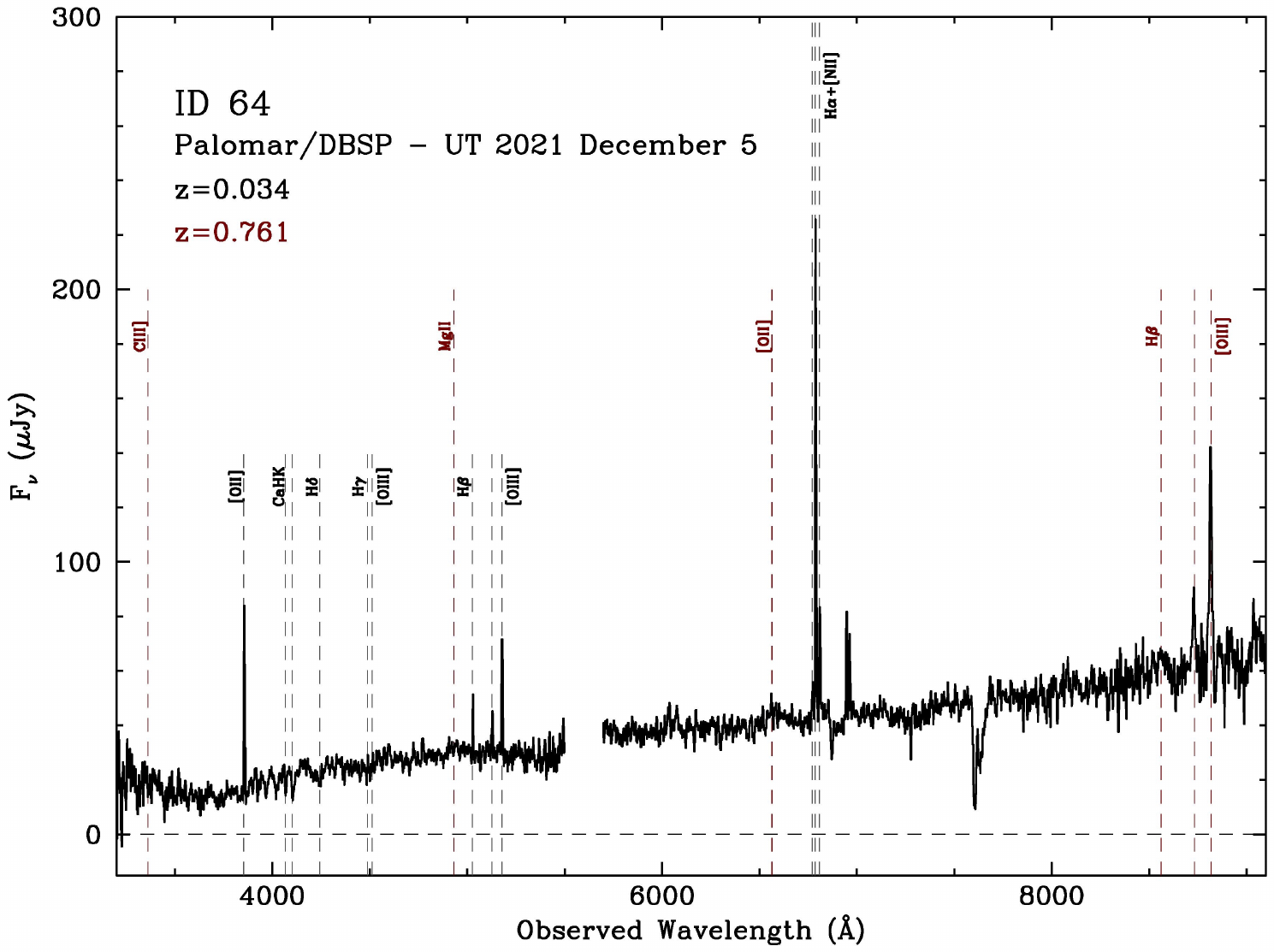}
    \caption{Palomar Double Spectrograph (DBSP) spectrum of the bright optical counterpart to the radio source in ID 64. Foreground narrow emission lines in the dwarf galaxy are marked in black and correspond to a redshift of $z=0.034$. There is another set of emission lines shown in red that correspond to a redshift of $z=0.761$ and that likely originate from a background AGN associated with the radio source.}
    \label{fig:spectrum64}
\end{figure*}

\section{Palomar Spectroscopy of ID 64}
\label{sec:spectra_ID64}

{The radio source in ID 64 is of particular interest since it is the only off-nuclear source in our sample that has a bright optical counterpart, and, therefore, we elected to take further observations of this object.} On the night of UT 2021 December 5, we obtained a single 1200~s exposure of ID 64 with the Double Spectrograph on the Palomar 5-m Hale Telescope. The night was photometric and the seeing was 1\farcs5 at the time of the observation.  The instrument was configured with a 1\farcs5 wide slit, which was oriented at a position angle of 109\degr\ east of north in order to simultaneously observe both the central region of the dwarf galaxy and the offset point source to the west that is spatially coincident with the radio source.

We processed the data with IRAF using standard techniques and flux-calibrated the spectrum using standard stars observed on the same night.  The reduced spectrum (shown in Figure \ref{fig:spectrum64}) reveals multiple narrow emission lines from the dwarf galaxy at $z = 0.034$, consistent with star formation.  Several of the emission lines are spatially extended, and we also detect absorption lines and a Balmer decrement consistent with star formation in the dwarf galaxy.  In addition, we detect two slightly broader lines redwards of 8500~\AA\ which are identified as the [\ion{O}{3}] doublet at $z = 0.761$.  The lack of associated H$\beta$ emission implies a high-ionization source with $\log$~[\ion{O}{3}]~5007/H$\beta \gtrsim 1$; i.e., an active galaxy according to the Baldwin, Phillips \& Terlevich (BPT) diagram \citep{Baldwin1981}. The lack of associated strong broad lines suggest that the interloper active galaxy is obscured.

\begin{figure*}
    \includegraphics[width=\textwidth]{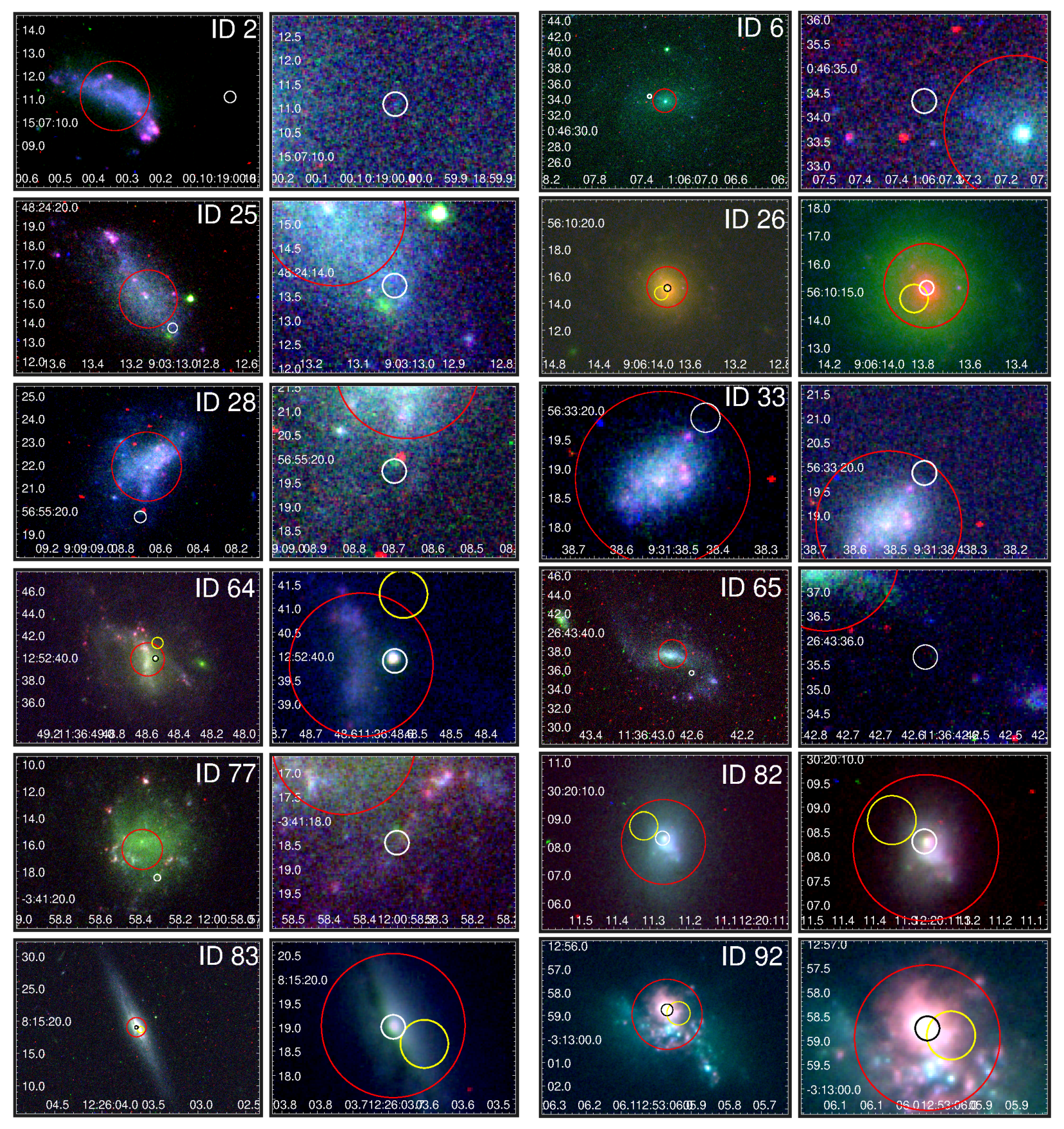}
    \caption{Three-color HST \: images of our galaxies. Green corresponds to the wide $I$-band F814W filter or wide $YJ$ F110W filter (for ID 26), red corresponds to the H$\alpha$ F680N or F665N filter (for ID 6) or wide $V$-band F606W filter (for ID 26) and blue corresponds to the wide $B$-band F475W or SDSS $g$-band F438W filter (for IDs 64, 77 and 92) or wide UV F275W filter (for ID 26). The white/black circles have radius 0\farcs25 and show the location of the compact radio source. The yellow circles show the location of the X-ray detections with radii {0\farcs5}. {The red circles indicate the location of the SDSS fibers with diameter of 3\farcs0.}  {IDs 26, 64, 82, 83 and 92 all have both optical counterparts to the radio sources observed in all filters and X-ray detections corresponding to roughly same sky location (also see Figure \ref{fig:xray_detections}).}
    }
    \label{fig:rgb}
\end{figure*}

\section{A Multi-wavelength View}
\label{sec:multiwavelength}

{In the initial radio study conducted by \citet{reines20}, through a detailed analysis calculating the expected emission at 9 GHz from various potential stellar sources within the dwarf galaxies (supernovae, supernovae remnants and thermal \HII \: regions), our 12 galaxies were found to have luminosities brighter than what any of these stellar sources could likely produce.\footnote{\citet{reines20} notes that, taking into account the uncertainties, the luminosity for ID 92 is consistent with an \HII \: region with an instantaneous star formation rate equivalent to that expected for the entire galaxy.} Therefore, \citet{reines20} conclude that these objects are very likely to be AGNs, either within the host dwarf galaxy or background objects. We combine this work at radio wavelengths with our new X-ray and optical data to further characterize the sources of the compact radio emission. }

\subsection{Radio Source Positions and their X-ray and Optical Counterparts}

Figure \ref{fig:rgb} shows {HST} three-color images for our sample of dwarf galaxies. The locations of the radio sources are marked with white or black circles with radii of 0\farcs25 (i.e., the resolution of the VLA observations). We also mark the positions of the X-ray sources for the five galaxies with detections using yellow circles with radii of 0\farcs5.

{We find five galaxies in our sample with significant X-ray and optical counterpart detections: IDs 26, 64, 82, 83 and 92. All five of these detections have X-ray source positions that are roughly consistent with the positions of the radio sources (see Figure \ref{fig:xray_detections}) and are unlikely to be background X-ray sources aligning with our target galaxies by chance \citep{moretti03, Flesch2023}. Besides the background AGN residing in ID 64 (see \S\ref{sec:spectra_ID64}), the rest of these galaxies have radio/optical/X-ray sources consistent with the center of the dwarf host galaxies. The remaining seven galaxies, which all have off-nuclear radio sources, do not have significant optical or X-ray counterparts. We discuss the implications of the optical and X-ray non-detections in Section \ref{sec:nondetections}. }

\subsection{X-ray Radio-Loudness}
\label{sec:xray_radio_loudness}

The X-ray radio-loudness parameter (R$_\mathrm{x}$) is defined as the ratio of radio luminosity (at 5 GHz) to the hard X-ray luminosity (from 2-10 kev), with log R$_\mathrm{x} > -4.5$ being considered ``radio-loud" \citep{terashima03, Ho2008}. We calculate this parameter for the five galaxies in our sample with overlapping X-ray and radio detections, assuming that the sources of the X-ray and radio emission are the same. We use the X-ray luminosities from \S\ref{sec:xraysrc} and estimate 5 GHz radio luminosities using the 9 GHz flux densities and spectral indices from \cite{reines20}. We find values in the range $-3.9 \leq$ log R$_\mathrm{x}$ $\leq -1.7$ and these objects are therefore considered radio-loud. We calculate uncertainties on the X-ray radio loudness parameter by propagating the errors on the measured X-ray fluxes and the errors reported by \citet{reines20} for S$_{9\mathrm{GHz}}$ and $\alpha$. Uncertainties for L$_{5\mathrm{GHz}}$ are $\sim$20\% and {uncertainties for R$_\mathrm{X}$ range from 4\%-30\% with one outlier (ID 25) at 60\%}. Lower limits on R$_\mathrm{x}$ are calculated for the radio sources without X-ray detections. The R$_\mathrm{x}$ values are given in Table \ref{tab:radio_loudness}.

{For comparison, the low-luminosity AGNs in the dwarf galaxies Henize 2-10 and Mrk 709 have log R$_\mathrm{x} = -2.4$ \citep{reines16} and $-3.7$ \citep{reines14}, respectively. The \citet{terashima03} sample of optically-selected, low-luminosity AGNs (LLAGNs), originally from the Palomar survey of bright objects \citep{Ho1995} and also observed in the radio \citep{Nagar2000}, have X-ray radio-loudness values in the range $-$4.6 $< \mathrm{R_X} < $ $-$1.2. The X-ray radio loudness values we find for our sample lie within this range.}
{Additionally, the low-mass AGNs studied in \citet{gultekin2022} have X-ray radio-loudness values in the range $-$5.3 $\lesssim \mathrm{R_X} \lesssim$ $-$1.4. The eight AGNs from \citet{gultekin2022} were originally selected as AGNs by \citet{reines13} based on their optical narrow emission line ratios and broad H$\alpha$ emission. These objects then had follow-up X-ray \citep{baldassare17} and radio observations \citep{gultekin2022}.} Given that the objects studied here were originally selected in the radio, it is not surprising that our sample generally has relatively high $\mathrm{R_X}$ values compared to optically-selected samples of AGNs.

\subsection{Optical Radio-Loudness}
\label{sec:optical_radio_loudness}

The optical radio-loudness parameter (R) describes the ratio of the luminosity density {(with units of erg s$^{-1}$ Hz$^{-1}$)} in the radio (at 5 GHz) to the optical \citep[4400 \AA,][]{kellermann89}, with R$>$10 being considered ``radio-loud." {Again, we estimate 5 GHz radio luminosities using the 9 GHz flux densities and spectral indices from \citet{reines20}}. We estimate the luminosity density at 4400 \AA\ through a linear interpolation in log space between the flux density measurements (or upper limits) in the shortest and longest wavelength filters. We report the radio-loudness parameter (and lower limits in the case of no detection in the optical images) for galaxies in our sample in Table \ref{tab:radio_loudness}. Our radio sources are almost all considered radio-loud, with only ID 92 having log R $<$ 1. {We estimate uncertainties on the optical radio loudness parameter by propagating the errors on the measured optical fluxes (see Section \ref{sec:optical_photometry}) and the errors reported by \citet{reines20} for S$_{9\mathrm{GHz}}$ and $\alpha$. Uncertainties for L(5GHz) are $\sim$20\% (as in Section \ref{sec:xray_radio_loudness}), uncertainties for L(4400 \AA) are $\sim$6\% and uncertainties for R are $\sim$15\%.}

We find that the optical radio-loudness values are within the typical range of known radio-loud quasars. We recalculate our radio-loudness parameters {(and corresponding uncertainties)} using the method described above, but at 2500 \AA \: for consistency with the work done by \cite{miller2011} and \citet[][, this parameter is referred to as R* in \citet{miller2011}, with R* $>$ 10 considered radio-loud]{Shen2011}. {\citet{Shen2011} reports values for all quasars from SDSS DR7 and \citet{miller2011} reports values for radio-loud quasars from SDSS DR5. Both samples were mostly optically-selected using the corresponding SDSS Quasar Catalogs \citep{schneider2007, schneider2010}. \citet{miller2011} also uses a photometric sample of optically-selected AGNs based on \citet{richards2009} classification and a ``quasi-radio-selected" sample, which consists of objects originally targeted by SDSS as potential optical counterparts to FIRST sources.} For both these samples, log R values range from $\sim$ $-$0.6 to 5, with our sample falling within this range. We note that extrapolating linearly in log-log space to find the flux density at 2500 \AA \: is less reliable than extrapolating to 4400 \AA\ given the filters used for our observations, as uncertainties for L (2500 \AA) are $\sim$7\% and uncertainties for R* are $\sim$20\%.
However, since we are calculating log R, the difference between our extrapolated flux density and the true value likely has a small impact.

\begin{deluxetable*}{ccccccc}
\tabletypesize{\footnotesize}
\tablecaption{Radio Loudness Parameters}
\tablewidth{0pt}
\tablehead{
\colhead{ID} &  \colhead{log L(5GHz)}  &  \colhead{log L(4400 \AA)} & \colhead{log L(2500 \AA)}   &  \colhead{log R} & \colhead{log R*} & \colhead{log R$_\mathrm{x}$} }
\startdata
2 & 21.91 & $<$18.11 & $<$17.82 & $>$3.80 & $>$4.10 & $>-$1.84 \\
6 & 20.44 & $<$17.68 & $<$17.35 & $>$2.77 & $>$3.10 & $>-$1.54 \\
25 & 21.07 & $<$18.25 & $<$17.95 & $>$2.82 & $>$3.12 & $>-$1.74 \\
26 & 21.86 & 19.91 & 19.25 & 1.95 & 2.62 & $-$2.14 \\
28 & 21.35 & $<$18.08 & $<$17.79 & $>$3.27 & $>$3.57 & $>-$1.84 \\
33 & 22.23 & $<$18.39 & $<$18.08 & $>$3.84 & $>$4.15 & $>-$1.14 \\
64\tablenotemark{\footnotesize{a}} & 21.84 & 19.25 & 19.00 & 2.60 & 2.84 & $-$3.94 \\
65 & 21.54 & $<$18.11 & $<$17.70 & $>$3.43 & $>$3.83 & $>-$2.14 \\
77 & 21.52 & $<$18.35 & $<$18.01 & $>$3.18 & $>$3.51 & $>-$1.64 \\
82 & 20.94 & 19.46 & 19.32 & 1.48 & 1.61 & $-$1.74 \\
83 & 21.11 & 19.27 & 18.77 & 1.84 & 2.34 & $-$2.74 \\
92 & 21.26 & 20.45 & 20.51 & 0.81 & 0.75 & $-$2.44 \\
\enddata
\tablecomments{Column 1: {Radio luminosity in W Hz $^{-1}$ at 5GHz calculated using the spectral index ($\alpha$) from \citet{reines20} and the relation $S_\nu \propto \nu^\alpha$}. Column 2-3: Optical luminosity in W Hz $^{-1}$ at 4400 \AA \: and 2500 \AA, respectively, interpolated from the {\it HST} flux measurements. For non-detections in relevant optical filters we report upper limits on the luminosity. Column 4-5: Optical radio-loudness parameters calculated at either 4400 \AA \: (R) or 2500 \AA \: (R*). log R (or log R*)  $>$ 1 is considered ``radio-loud." For non-detections in relevant optical filters we report lower limits on log R. Column 6: X-ray radio-loudness parameter. log R$_\mathrm{x}$   $>$ -4.5 is considered ``radio-loud". We use 2-10 keV X-ray luminosities listed in Table \ref{tab:xray}. X-ray non-detections we report lower limits on log R$_\mathrm{x}$. {See text in relevant sections for uncertainties on these values.} }
\tablenotetext{a}{{The luminosities reported for this object are based on the distance of the dwarf galaxy, not the redshift of the background quasar.}}
\label{tab:radio_loudness}
\end{deluxetable*}

\subsection{Implications of Optical/X-ray Non-detections}
\label{sec:nondetections}

Seven galaxies in our sample do not have significantly detected optical counterparts in the {HST} imaging at the locations of the radio sources, all of which are spatially offset from the centers of the dwarf galaxies. Here we consider the implications of these non-detections.

If the radio-detected AGNs are wandering BHs in the dwarf host galaxies, they may be isolated BHs without a stellar counterpart, {possibly accreting from either the ISM or a retained accretion disk \citep{Bonning2007, Blecha2011},} and we would not detect an optical source. Alternatively, the BH may be surrounded by a stellar cluster that we could potentially detect. We use our derived upper limits in the {HST} imaging to constrain the properties of a putative detectable star cluster. We model a star cluster in each galaxy using the GALEV simple stellar population (SSP) synthesis model \citep{Kotulla2009}, with Padova isochrones, a timestep of 4 Myr, a Kroupa initial mass function (IMF) spanning 0.1 - 100 \msun, and a metallicity of [Fe/H] = $-$0.7. We take this model at ages 10 Myr, 100 Myr, and 1 Gyr and scale the spectra given the redshift of each galaxy. We convolve this GALEV model with the F814W filter throughput curve to determine a model flux density in this band. Assuming the flux density scales linearly with the mass of the star cluster, we calculate the upper limit {range\footnote{Ranges are given for the upper limits since the  distances to galaxies in the analyzed sample vary significantly (between 70 and 203 Mpc).}} on the possible mass of a cluster that we would not be able to detect at optical wavelengths. We find that {a cluster at 10 Myr would be below our detection limits if it had log $(M/M_\odot) \lesssim$ 5.0-5.7, while the older clusters could be even larger with log $(M/M_\odot) \lesssim$ 5.8-6.5 and log $(M/M_\odot) \lesssim$ 6.3-7.0, for 100 Myr old and 1 Gyr old clusters, respectively}. These are typical masses for globular and nuclear star clusters, and, therefore, it remains a possibility that there are star clusters hosting wandering BHs in these dwarf galaxies that are below our detection limits. {With deeper {\it JWST} observations, we would ideally be able to observe the potential star clusters accompanying these radio sources.}

{Additionally, as exemplified by ID 64 (\S \ref{sec:spectra_ID64}), it is possible that these radio sources are not associated with the dwarf galaxies at all and instead are background interlopers that happen to overlap with our galaxies in the sky. We consider possible galaxy hosts at various redshifts to derive upper luminosity limits on potential AGN host galaxies that would be below our optical detection limits. We use the flux density limits for our non-detections and calculate the luminosity in the F814W filter {(roughly I-band)} at various redshifts. {We find that the putative host galaxies would have log$(L/L_\odot) \lesssim$ {9.1, 9.8, 10.6, and 11.0} for $z = 0.5, 1, 2$ and 3, respectively. }

We now consider the X-ray non-detections. {Seven} of the radio sources do not have detectable X-ray counterparts. These sources tend to have relatively high upper limits on their X-ray radio loudness parameters (see Table \ref{tab:radio_loudness}){, i.e., they are under-luminous in the X-ray band}. {One potential explanation is that some of the AGNs in these objects{, e.g., IDs 6 and 65,} are highly obscured in the X-ray band. For example, if ID 6 was obscured by a column density of $\gtrsim 10^{24}\,\mathrm{cm}^{-2}$, it could hide an AGN with $\log(L_{2-10\,\mathrm{kev}})\sim 40.7$ erg s$^{-1}$.  It is also possible that the AGN emission has changed, as the radio and X-ray observations are non-simultaneous{; in this case, the sources could be “changing-look” AGNs, exhibiting luminosity variations of roughly an order of magnitude over timescales of years \citep[e.g.,][]{Ricci2020}}.}} Finally, AGNs in dwarf galaxies may suffer from intrinsic X-ray weakness \citep[e.g.,][]{arcodia2024}. {The {Chandra} X-ray observations obtained as part of this program are fairly shallow. Deeper observations would allow the placement of firmer constraints on the non-detections and, in the cases of detected sources, much better X-ray spectra could provide more insights into the nature of these objects, including their X-ray time variability.}

\subsection{Notes on Individual Objects}
\label{sec:individual objects}

Each galaxy in our sample has a compact radio source with a luminosity strongly suggesting the presence of an AGN \citep{reines20}. In this section, we aim to use multi-wavelength observations to help determine if a given AGN resides in its apparent dwarf galaxy host or is a background AGN. In addition to the {Chandra} and {HST} observations presented here, we also include information from SDSS spectroscopy (taken from \citealt{reines20}) and the VLBA observations of \citet{Sargent2022} {at 9 GHz with milliarcsecond resolution. Due to the smaller beam size of VLBA versus VLA, nondetections in the VLBA observations indicate more extended radio emission and lower brightness temperatures}. Additional information from the literature is also considered. As discussed above, all the radio-detected AGNs are considered radio-loud compared to their luminosities (or upper limits) at X-ray and optical wavelengths (with the exception of ID 92 in the optical). Table \ref{tab:summary} summarizes the multi-wavelength results.

\textbf{ID 2}: This radio source is significantly offset from the optical center of the galaxy (4\farcs9). It has neither an X-ray nor an optical counterpart detected. The galaxy is classified as star-forming in all three narrow-line diagnostic diagrams based on its SDSS optical spectrum. However, the SDSS fiber does not overlap with the radio source. The VLA radio source is detected with the VLBA on parsec scales, and its distance from the photocenter of the dwarf galaxy is consistent with expectations for a background AGN. 

\textbf{ID 6}: This radio source is off-nuclear, lying 2\farcs1 from the optical center of the galaxy. The galaxy is classified as star-forming in the \OIII/\hbeta\ versus \NII/\ha\ and \SII/\ha\ diagrams and as a LINER in the \OIII/\hbeta\ versus \OI/\ha\ diagram. However, the SDSS fiber does not overlap with the radio source so these classifications should be taken with caution. The VLA radio source does not have detectable VLBA, X-ray, or optical counterparts and we cannot distinguish between a wandering BH in the dwarf galaxy or a background AGN.

\textbf{ID 25}: This radio source is off-nuclear, lying 2\farcs2 from the optical center of the galaxy. The galaxy is classified as star-forming in the \OIII/\hbeta\ versus \NII/\ha\ and \SII/\ha\ diagrams and it is classified as a Seyfert in the \OIII/\hbeta\ versus \OI/\ha\ diagram. However, the SDSS fiber does not overlap with the radio source. The VLA radio source has no VLBA, X-ray nor optical counterpart detected and we cannot distinguish between a wandering BH in the dwarf galaxy or a background AGN.  

\textbf{ID 26}: This radio source is an AGN that has a position consistent with the nucleus of the dwarf galaxy. It has both an X-ray and optical counterpart detected in the same location as the radio source. This is the only galaxy in our sample that is classified as an AGN/Seyfert in all three optical narrow emission line ratio diagrams and the SDSS fiber overlaps with the radio, X-ray and optical sources. This object was previously confirmed to be an AGN based on optical emission line ratios and broad H$\alpha$ emission \citep[RGG 9 in][]{reines13}, as well as mid-IR colors \citep[ID 2 in][]{latimer21}. It was not detected by the VLBA. 

{This galaxy was previously studied in X-rays by \citet{baldassare17}, where they found that the 2-10 keV X-ray luminosity versus the galaxy SFR (calculated using \ha \: emission measured using a 3\farcs0 SDSS fiber) exceeded expectations from XRBs and indicated the presence of an AGN. In our work, while we measure a very similar 2-10 keV X-ray luminosity{, we also investigate the X-ray spectrum of the galaxy and find that it points towards a SF origin for the X-ray emission or at least a significant SFR component. However, we urge caution here since the X-ray spectrum is of poor quality with a low number of counts.}

\begin{deluxetable*}{cccccccccccc}
\tabletypesize{\footnotesize}
\tablecaption{Detection Summary}
\tablewidth{0pt}
\tablehead{
\colhead{ID} & \colhead{Position} & \multicolumn{4}{c}{SDSS Spectra} & \multicolumn{2}{c}{\textit{Chandra} X-ray} & \colhead{\textit{HST} Optical} & \colhead{Previous Work} \\
\cmidrule(l){3-6} \cmidrule(l){7-8} \cmidrule(l){9-9}
\colhead{ } &  \colhead{ } & \colhead{\NII/\ha} & \colhead{\SII/\ha} & \colhead{\OI/\ha} & \colhead{Overlap} & \colhead{Detected} & \colhead {AGN/ULX} & \colhead{Detected} & \colhead{ } \\ 
\colhead{ (1)}  & \colhead{ (2)} & \colhead{ (3)} & \colhead{(4)} & \colhead{(5)} & \colhead{(6)} & \colhead{ (7)} & \colhead{(8)} & \colhead{(9)} & \colhead{(10) } } 
\startdata
2 & Wandering  & SF  & SF      & SF      & No  & - & -          & - & VLBA Bkg AGN  \\
6 & Wandering  & SF  & SF      & LINER   & Yes & - & -          & - & -             \\
25 & Wandering & SF  & SF      & Seyfert & No  & - & -          & - & -             \\
26 & Nuclear   & AGN & Seyfert & Seyfert & Yes & X & AGN & X & Confirmed AGN \\
28 & Wandering & SF  & SF      & SF      & No  & - & -          & - & VLBA Bkg AGN  \\
33 & Wandering & SF  & SF      & SF      & Yes & - & -          & - & -             \\
64 & Wandering & SF  & SF      & Seyfert & Yes & X & AGN   & X & -             \\
65 & Wandering & SF  & LINER   & Seyfert & No  & - & -   & - & VLBA Bkg AGN  \\
77 & Wandering & SF  & SF      & SF      & No  & - & -          & - & -             \\
82 & Nuclear   & SF  & SF      & SF      & Yes & X & AGN          & X & Confirmed AGN \\
83 & Nuclear   & SF  & SF      & Seyfert & Yes & X & AGN   & X & -             \\
92\tablenotemark{\footnotesize{a}} & Nuclear   & -   & -       & -       & -   & X & ULX & X & -              \\
\enddata
\tablecomments{Column 1: Identification number.
Column 2: Whether the radio source is considered wandering or nuclear.
Column 3: \OIII/\hbeta \: versus \NII/\ha \: classification based on SDSS spectroscopy.
Column 4: \OIII/\hbeta \: versus \SII/\ha \: classification based on SDSS spectroscopy.
Column 5: \OIII/\hbeta \: versus \OI/\ha \: classification based on SDSS spectroscopy.
Column 6: Whether or not the SDSS fiber overlaps with the radio source.
Column 7: Whether or not there is an X-ray detection at any location in the galaxy.
Column 8: Likely origin of X-ray emission (see Sec \ref{sec:xrbs}).
Column 9: Whether or not there is an optical counterpart to the radio source in all optical filters.
Column 10: Any previous results about the radio source. \citet{Sargent2022} provides VLBA data. \citet{reines13} and \citet{baldassare17} confirm the identity of ID 26. \citet{molina21} confirms the identity of ID 82.
\tablenotetext{a}{This galaxy's SDSS spectrum was unreliable at \ha \: so we cannot classify this galaxy based on narrow line diagnostics.} 
}
\label{tab:summary}
\end{deluxetable*}

\textbf{ID 28}: This radio source is offset by  2\farcs7 from the optical center of the galaxy. The galaxy is classified as star-forming in all three optical narrow emission line ratio diagrams based on its SDSS optical spectrum. However, the SDSS fiber does not overlap with the radio source. The radio source does not have a detected X-ray or optical counterpart. This object was detected by the VLBA on parsec scales, and its distance from the center of the dwarf galaxy is consistent with expectations for a background AGN.

\textbf{ID 33}: This radio source is off-nuclear, lying 1\farcs1 from the optical center of the galaxy. It is classified as a star-forming galaxy based on all three optical narrow emission line ratio diagrams based on its SDSS optical spectrum. The radio source is positioned near the edge of the SDSS fiber and the fiber also encompasses the entirety of the dwarf galaxy. The VLA radio source is not detected in X-rays or in the optical, nor is it detected with the VLBA. We cannot distinguish between a wandering BH in the dwarf galaxy or a background AGN.

\textbf{ID 64}: This radio source is off-nuclear, lying 0\farcs9 from the optical center of the galaxy. {We detect both a bright {HST} optical and {Chandra} X-ray counterpart. Our Palomar spectrum reveals that, in addition to the narrow emission lines originating from the dwarf galaxy at redshift $z=0.034$, there is also a second component corresponding to an AGN at higher redshift ($z=0.761$). This strongly suggests the radio source is due to a background AGN. Given this redshift and our measured fluxes, we calculate the following luminosities: $\mathrm{L_{9GHz}} = 2.6 \times 10^{39}$ erg s$^{-1}$, $\mathrm{L_{2-10 \: keV}} = 7.4 \times 10^{44}$ erg s$^{-1}$ and $\mathrm{L_{optical}} \sim 10^{40}$ erg s$^{-1}$. While the galaxy is classified as star-forming in the \OIII/\hbeta\ versus \NII/\ha\ and \OIII/\hbeta \: versus \SII/\ha \: diagrams, it is classified as a Seyfert in the \OIII/\hbeta \: versus \OI/\ha\ diagram based on the SDSS optical spectrum. This background AGN is not detected with the VLBA. }
{However, this background AGN may very well be the detected X-ray source, as its spectrum is consistent with that of a Seyfert-type AGN. }

\textbf{ID 65}: This radio source is positioned 2\farcs9 from the optical center of the galaxy. The galaxy is classified as star-forming in the \OIII/\hbeta\ versus \NII/\ha\ diagram, as a LINER in the \OIII/\hbeta\ versus \SII/\ha\ diagram, and as a Seyfert in the \OIII/\hbeta\ versus \OI/\ha\ diagram. However, the SDSS fiber does not overlap with the radio source so these classifications should be taken with caution. The radio source does not have a detected X-ray or optical counterpart. It is detected with the VLBA on parsec scales and is consistent with expectations for a background AGN. Alternatively, \citet{Dong2024} suggest this object could be a persistent radio source (PRS) associated with a fast radio burst (FRB) based on its compact size and host-normalized offset.

\textbf{ID 77}: This radio source is offset, lying 2\farcs2 from the optical center of the galaxy. It is classified as a star-forming galaxy in all three optical narrow emission line ratio diagrams based on the SDSS optical spectra. However, the SDSS fiber does not overlap with the radio source. The VLA radio source has neither an X-ray nor an optical counterpart detected, nor is it detected with the VLBA. We cannot distinguish between a wandering BH in the dwarf galaxy or a background AGN.

\textbf{ID 82}: This radio source has a position consistent with the nucleus of the dwarf galaxy. Both an X-ray and optical counterpart were detected for this radio source, but it was not detected with the VLBA. {This object was only detected in the hard X-ray band and is consistent with an obscured AGN.} It is classified as a star-forming galaxy in all three optical narrow emission line ratio diagrams based on the SDSS optical spectrum, and the SDSS fiber overlaps with the radio source. Despite the SDSS spectral classification of star-forming and the VLBA non-detection, this object was confirmed to be an AGN by the coronal [Fe X] line, enhanced \OI \: emission at the same location of the radio source, and broad H$\alpha$ emission \citep{molina21}. 

\textbf{ID 83}: This radio source is located in the nucleus of the dwarf galaxy. It has both an X-ray and optical counterpart detected at a sky location consistent with the position of the VLA radio source, but no detection with the VLBA. {Its X-ray luminosity and spectrum are consistent with those of an AGN.} It is classified as a Seyfert galaxy in the \OIII/\hbeta\ versus \OI/\ha\ diagram\ (and star-forming in the other two diagrams) and the SDSS fiber overlaps with the radio source. There are multiple lines of evidence indicating this galaxy hosts a central AGN.

\textbf{ID 92}: This radio source is co-located with a giant \HII \: region that is visible in the F680N narrowband image. While the SDSS fiber overlaps with the radio source, the spectrum is unreliable at \ha\ and so we cannot definitively classify this galaxy based on narrow line diagnostics. The VLA radio source is not detected with the VLBA. There is an X-ray detection consistent with the position of the radio source, though there are many other optical sources (i.e., star clusters) that could be counterparts of the X-ray source and the X-ray luminosity is consistent with that of a ULX given the SFR of the galaxy. This object is our only target that is considered radio-quiet relative to its optical luminosity. \citet{reines20} also pointed out that the radio source was consistent with an \HII \: region within their uncertainties (see their Figure 6).

Here we leverage our high-resolution {HST} imaging to revisit the origin of this radio source. {The production rate of Lyman continuum photons ($Q_\mathrm{Lyc}$) can be calculated both from the radio emission (assuming a thermal \HII\ region) and from the \ha\ emission. By comparing these $Q_\mathrm{Lyc}$ values, we can assess whether or not the radio-based value is greatly in excess of the \ha-based value.} While some difference could be attributed to extinction \citep{reines08}, {a radio-based value more than $\sim$an order of magnitude larger} could indicate the presence of an AGN boosting $Q_\mathrm{Lyc}$.

We first use the following equation from \citet{condon92}:

\begin{equation}
    \begin{split}
        \left( \frac{Q_\mathrm{Lyc}}{\mathrm{s^{-1}}} \right) \gtrsim 6.3 \times 10^{52} \left( \frac{\mathrm{T_e}}{10^4 \mathrm{K}} \right)^{-0.45} \left( \frac{\nu}{\mathrm{GHz}} \right)^{0.1} \\ 
        \times \left( \mathrm{\frac{L_{\nu,\mathrm{thermal}}}{10^{27} erg \: s^{-1} Hz^{-1}}} \right),
    \end{split}
\end{equation}

\noindent 
(with $\mathrm{T_e = 10^4~K}$) along with the radio spectral luminosity at 9 GHz from \citet{reines20} to predict the radio-based estimate of $Q_\mathrm{Lyc}$.

Then we use the equation below along with the relation between the flux of \ha\ and \hbeta\ assuming Case B recombination, $F_{H\alpha}/F_{H\beta} = 2.73$ \citep{condon92}, to again predict $Q_\mathrm{Lyc}$, but this time using the {HST}-measured \ha\ flux:

\begin{equation}
        \left( \frac{Q_\mathrm{Lyc}}{\mathrm{s^{-1}}} \right) \gtrsim 2.25 \times 10^{12} \left( \frac{\mathrm{T_e}}{10^4 \mathrm{K}} \right)^{0.07} \left( \mathrm{\frac{L_{H\beta}}{erg \: s^{-1}}} \right).
\end{equation}

\noindent 
We calculate $F_{H\alpha}$ using our HST \: imaging as:

\begin{equation}
    F_{H\alpha} = [f_{\mathrm{F680N}} - f^{(H\alpha)}_\mathrm{cont}] \Delta \lambda_{\mathrm{F680N}}
\end{equation}

\noindent 
where we estimate $f^{(\rm H\alpha)}_\mathrm{cont}$ by interpolating between the two wide band filters in log space and $\Delta \lambda_{\mathrm{F680N}}$ is the width of the narrow-band filter.

{We find that the radio value is somewhat higher than the \ha-based value} (by a factor of 5): log $Q_\mathrm{Lyc, radio} = 54.0 \: \mathrm{s}^{-1}$ and log $Q_{\rm Lyc, optical} = 53.3 \: \mathrm{s}^{-1}$. {The difference between the Lyman continuum flux estimates does not necessarily warrant an interpretation of an AGN origin for the radio emission.} We also note that the circular aperture in which the \ha\ emission is measured ($r=0\farcs2$) has a similar area as the radio source size ($0\farcs24 \times 0\farcs2$). 
{From these $Q_\mathrm{Lyc}$ values and the relation from \citet{kennicutt98}, $\mathrm{SFR (M_\odot \: yr^{-1}) = 1.08 \times 10^{-53} \: Q_{Lyc} \: (s^{-1})}$, we calculate the instantaneous star formation rate surface density as $\sim$100 M$_\odot$ yr$^{-1}$ kpc$^{-2}$ using $Q_\mathrm{Lyc, optical}$ and $\sim$400 M$_\odot$ yr$^{-1}$ kpc$^{-2}$ using $Q_\mathrm{Lyc, radio}$. While these exceed the maximum starburst intensity limit from \citet{meurer1997} calculated from UV, \ha, far-IR and radio continuum emission over a wide redshift range, these values are not completely unprecedented. \citet{Diamond-Stanic2012} study three galaxies with star formation rate surface densities of $\sim 3000$ M$_\odot$ yr$^{-1}$ kpc$^{-2}$. They conclude that these high values are not due to contamination from AGNs and are likely from a short-lived phase of compact, extreme star-formation.

We calculate the implied extinction by comparing the measured \ha\ flux ($F_{\rm H\alpha, observed} = 2.11 \times 10^{-13}$ erg s$^{-1}$ cm$^{-2}$) to that predicted from the radio continuum \citep{condon92}:

\begin{equation}
    \begin{split}
        \left( \frac{F_\mathrm{H\alpha,\mathrm{predicted}}}{\mathrm{erg\:  s^{-1} cm^{-2}}} \right) \sim 0.8 \times 10^{-12} \left( \frac{\mathrm{T_e}}{10^4 \mathrm{K}} \right)^{-0.59} \\ 
         \times \left( \frac{\nu}{\mathrm{GHz}} \right)^{0.1} \left( \mathrm{\frac{S_{\nu,\mathrm{thermal}}}{\mathrm{mJy}}} \right)
    \end{split}
\end{equation}

\noindent
{Using this equation with S$_{9 \mathrm{ GHz}}$=1.16 mJy from \citet{reines20} and assuming T$_\mathrm{e}=10^4$~K, we calculate $F_{\rm H\alpha, predicted} = 1.16 \times 10^{-12}$ erg s$^{-1}$ cm$^{-2}$.} The extinction at \ha\ is then given by

\begin{equation}
    A_{\rm H\alpha} = -2.5 \times {\rm log} \left( \frac{F_{\rm H\alpha, observed}} {F_{\rm H\alpha, predicted}}\right)
\end{equation}

\noindent
and we find $A_{\rm H\alpha}$ = 1.85 mag. This is within the range of known extragalactic massive star forming regions \citep{reines08} and therefore an AGN is not required to explain the radio luminosity.

\section{Conclusions}
\label{sec:conclusion}

We have presented {Chandra} and {HST} observations of 12 dwarf galaxies with radio-selected AGNs from \citet{reines20} to add to a growing body of follow-up observations \citep{Sargent2022, Dong2024} of this sample of (sometimes wandering) massive BHs. We use a series of multi-wavelength analyses to characterize the sources of emission from these galaxies. Our primary findings are summarized below.

\begin{enumerate}

    \item We detect both X-ray and optical counterparts of five radio sources (4 nuclear, 1 off-nuclear), with non-detections for the remaining seven. X-ray luminosities are in the range $L_{\rm 2-10~keV} \sim 10^{40-42}$ erg s$^{-1}$ assuming the sources are in the target dwarf galaxies.

    \item {Of the 4 nuclear radio sources with X-ray/optical detections, we find compelling multi-wavelength evidence for the presence of AGNs in three of our target dwarf galaxies: IDs 26, 82 and 83. We determine that the other dwarf galaxy may not host a radio AGN at all, as its properties are consistent with an extreme starburst region (ID 92). This object barely made the AGN sample of \citet{reines20} and our reanalysis incorporating our {HST} \ha\ imaging confirms the radio source is consistent with a giant \HII\ region with an extremely high star formation rate surface density ($\sim$100 M$_\odot$ yr$^{-1}$ kpc$^{-2}$). The X-ray source in ID 92 is consistent with a ULX.
}

    \item {We obtained a Palomar DBSP spectrum of the bright optical counterpart of the off-nuclear radio source towards ID 64 and identify it as a background AGN at z=0.761. The X-ray positional uncertainties prevent us from definitively saying whether or not the X-ray source originates from the background AGN or the target dwarf galaxy.}
    
    \item {Three of the galaxies in our sample have radio AGNs that are detected by the VLBA at parsec scales: IDs 2, 28, and 65. Nevertheless, none have optical or X-ray counterparts. These three radio sources are the most distant from the photocenters of the dwarf galaxies in our sample. Given the interloper modeling presented by \citet{reines20} and \citet{Sargent2022}, it is likely that these sources are background AGNs and are unrelated to their target dwarf galaxies.}

    \item Four other galaxies in our sample have offset radio AGNs. However, we cannot distinguish between wandering massive BHs in the target dwarf galaxies and background AGNs for these objects (IDs 6, 25, 33, and 77). None of them have detected X-ray or optical counterparts, nor are they detected at radio wavelengths on parsec scales with the VLBA \citep{Sargent2022}.

     \item While we do not detect optical counterparts for all but one (ID 64) of the off-nuclear radio sources, we find that the possible hosts could simply be below our {HST} detection limits. In the case of wandering BHs, they may be hosted by stellar clusters fainter than our sensitivity limits. For example, we would not detect a 100 Myr-old cluster with a stellar mass below $\sim 10^{6.2} M_\odot$.

\end{enumerate}

{Identifying the origin of the off-nuclear radio sources towards these galaxies may only be possible using the exquisite capabilities of the {James Webb Space Telescope (JWST)}. Sensitive, high-resolution observations with {JWST} would allow us to distinguish between any potential stellar overdensities (e.g., the core of a disrupted dwarf galaxy or a star cluster) affiliated with the radio source and residing within the dwarf galaxy (at $z\approx$ 0) from a background galaxy/AGN at higher redshift. If the presence of ``wandering BHs" is verified, it would confirm predictions from simulations that massive BHs need not always reside in the nuclei of dwarf galaxies \citep{bellovary19} and have important consequences for constraining BH seed formation using dwarf galaxies.

\vspace{1cm}

{We thank Lilikoi J. Latimer for her preliminary work on the X-ray analysis in this paper and the referee for helpful comments.} Support for this work was provided by NASA through Chandra Award No.\ GO2-23076X issued by the Chandra X-ray Observatory Center, which is operated by the Smithsonian Astrophysical Observatory for and on behalf of the NASA under contract NAS8-03060. Based on observations with the NASA/ESA Hubble Space Telescope obtained from MAST at the Space Telescope Science Institute, which is operated by the Association of Universities for Research in Astronomy, Incorporated, under NASA contract NAS5-26555. Support for Program numbers HST-GO-16843.001-A and HST-GO-16661.001-A were provided through grants from the STScI under NASA contract NAS5-26555. AER acknowledges support provided by NASA through EPSCoR grant number 80NSSC20M0231 and the NSF through CAREER award 2235277. The work of DS and portions of the work of TC were carried out at the Jet Propulsion Laboratory, California Institute of Technology, under a contract with NASA. {AB, RK and TC acknowledge support from the Smithsonian Institution and the Chandra Project through NASA contract NAS8-03060.} The National Radio Astronomy Observatory is a facility of the National Science Foundation operated under cooperative agreement by Associated Universities, Inc.

\bibliography{ref}
\appendix

\section{X-ray spectra} \label{sec:appendix}
\begin{figure}[h]
    \centering
    \includegraphics[width=\linewidth]{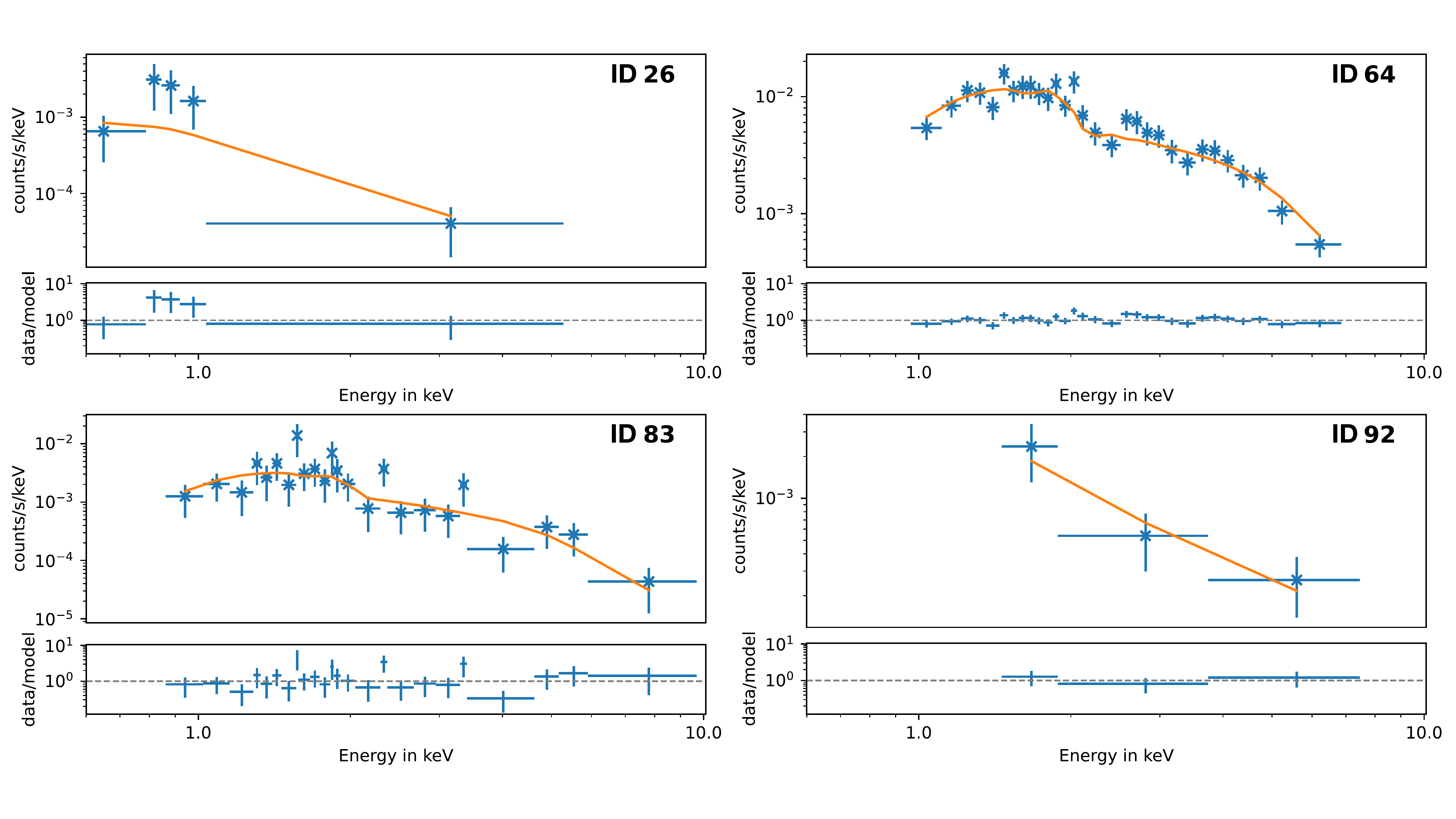}
    \caption{Observed {Chandra} X-ray spectra with sufficient counts for spectral modeling, best fit power law model (orange line) and ratio residuals. The data have been rebinned for plotting. The simple power law model with fixed Galactic absorption describes IDs 64, 83, and 92 well, but the spectral shape of ID 26 hints at the presence of a feature at $\sim 0.9$\,keV.}
    \label{fig:xrayspectra}
\end{figure}
\end{document}